\documentclass[journal]{sty/IEEEtran}

\usepackage[colorinlistoftodos,prependcaption,textsize=small]{todonotes}
\usepackage{color}
\usepackage{amsmath,amssymb,graphicx,booktabs,url}
\usepackage[utf8]{inputenc}
\usepackage{microtype}
\usepackage{stackengine}
\usepackage{tikz}
\usetikzlibrary{shapes.multipart}
\usepackage{multirow}
\usepackage{cite}
\usepackage{cleveref,xcolor,colortbl}
\usepackage{subcaption}

\title{On Loss Functions for Supervised Monaural\\ Time-Domain Speech Enhancement}

\author{\IEEEauthorblockN{Morten Kolbæk,
		Zheng-Hua Tan, \IEEEmembership{Senior Member, IEEE}, Søren Holdt Jensen, and
		Jesper Jensen}

\thanks{Manuscript received August 27, 2019; revised December 19, 2019; accepted January 18, 2020. The associate editor coordinating the review of this manuscript and approving it for publication was Prof. Tan Lee. \;\;\;\;\;\;\;\;\;\; \emph{Corresponding author: Morten Kolbæk}}%
\thanks{M. Kolbæk, Z.-H. Tan, and S. H. Jensen are with the Department of Electronic Systems, Aalborg University, Aalborg 9220, Denmark (e-mail: mok@es.aau.dk; zt@es.aau.dk; shj@es.aau.dk).}%
\thanks{J. Jensen is with the Department of Electronic Systems, Aalborg University, Aalborg 9220, Denmark, and also with Oticon A/S, Smørum 2765, Denmark (e-mail: jje@es.aau.dk).}%
\thanks{Digital Object Identifier 10.1109/TASLP.2020.2968738}	
}

\begin{document}

\newcommand\barbelow[1]{\stackunder[1.2pt]{$#1$}{\rule{.8ex}{.075ex}}}
\newcommand{\sd}{\barbelow{{d}}}
\newcommand{\s}{\barbelow{{a}}}
\newcommand{\one}{\barbelow{{1}}}
\newcommand{\sh}{\barbelow{{\hat{a}}}}


\maketitle
\begin{abstract}
Many deep learning-based speech enhancement algorithms are designed to minimize the mean-square error\,(MSE) in some transform domain between a predicted and a target speech signal. However, optimizing for MSE does not necessarily guarantee high speech quality or intelligibility, which is the ultimate goal of many speech enhancement algorithms. 
Additionally, only little is known about the impact of the loss function on the emerging class of time-domain deep learning-based speech enhancement systems. 

We study how popular loss functions influence the performance of time-domain deep learning-based speech enhancement systems.
First, we demonstrate that perceptually inspired loss functions might be advantageous over classical loss functions like MSE. 
Furthermore, we show that the learning rate is a crucial design parameter even for adaptive gradient-based optimizers, which has been generally overlooked in the literature.
Also, we found that waveform matching performance metrics must be used with caution as they in certain situations can fail completely. 
Finally, we show that a loss function based on scale-invariant signal-to-distortion ratio (SI-SDR) achieves good general performance across a range of popular speech enhancement evaluation metrics, which suggests that SI-SDR is a good candidate as a general-purpose loss function for speech enhancement systems.   

\end{abstract}

\section{Introduction}
\label{sec:intro}
Speech enhancement algorithms for improving speech quality and speech intelligibility of single-channel recordings of noisy speech are of high demand in a wide range of applications e.g. hearing aids design, mobile communications devices, voice-operated human-machine interfaces, etc.       
Consequently, developing successful monaural speech enhancement algorithms has been a long-lasting goal in both academia and industry. 

In fact, over the last decade, monaural speech enhancement algorithms based on machine learning, and deep learning in particular, have received a tremendous amount of attention (see e.g. \cite{kim_algorithm_2009,han_classification_2012,wang_towards_2013,xu_experimental_2014,weninger_single-channel_2014,healy_algorithm_2015,chen_large-scale_2016,erdogan_deep_2017,kolbaek_speech_2017,kolbaek_relationship_2019} as well as \cite{wang_supervised_2018,kolbaek_single-microphone_2018} and references therein).
Specifically, in recent years, deep learning-based speech enhancement algorithms, facilitated by powerful general-purpose graphics processing units and large amounts of training data, have shown impressive results by improving speech intelligibility in narrow acoustical conditions   \cite{healy_algorithm_2015,healy_algorithm_2017,bolner_speech_2016,monaghan_auditory_2017,goehring_speech_2017,lai_deep_2017,lai_deep_2018,healy_deep_2019}.   
However, despite the recent success of deep learning-based speech enhancement algorithms, many of the techniques referenced above are fundamentally limited, as they primarily focus on enhancement in the short-time spectral amplitude\,(STSA) domain and therefore ignore potentially useful phase information.
Numerous recent deep learning-based speech enhancement techniques exist, however, that incorporate phase information (e.g. \cite{roux_phasebook:_2019,wang_deep_2019,wang_end--end_2018}).
The most successful approaches to date are arguably end-to-end techniques based on fully convolutional neural networks\,(FCNN) that do not apply the short-time discrete Fourier transform\,(STFT) or other pre-processing stages, but operate directly in the time-domain (e.g. \cite{pandey_new_2018,pandey_new_2019,fu_end--end_2018,park_fully_2017,fu_raw_2017,pandey_tcnn:_2019,grzywalski_using_2019,tan_real-time_2019}).
These techniques, however, might still be limited as most of them rely on a loss function based on the mean square error\,(MSE) between time-domain waveforms. 
This is most likely suboptimal with respect to speech quality and intelligibility, as time-domain MSE has no apparent relation to human perception or the human auditory system in general.
Furthermore, as the works above use widely different network architectures, development datasets, noise types, hyperparameters, etc., it is not yet established how the loss functions influence the performance of such systems and if alternative loss functions that are more perceptually meaningful might be advantageous.

\begin{figure*}[ht]
	\centering
	\includegraphics[trim={0mm 0mm 0mm 0mm},clip,width=1.0\linewidth]{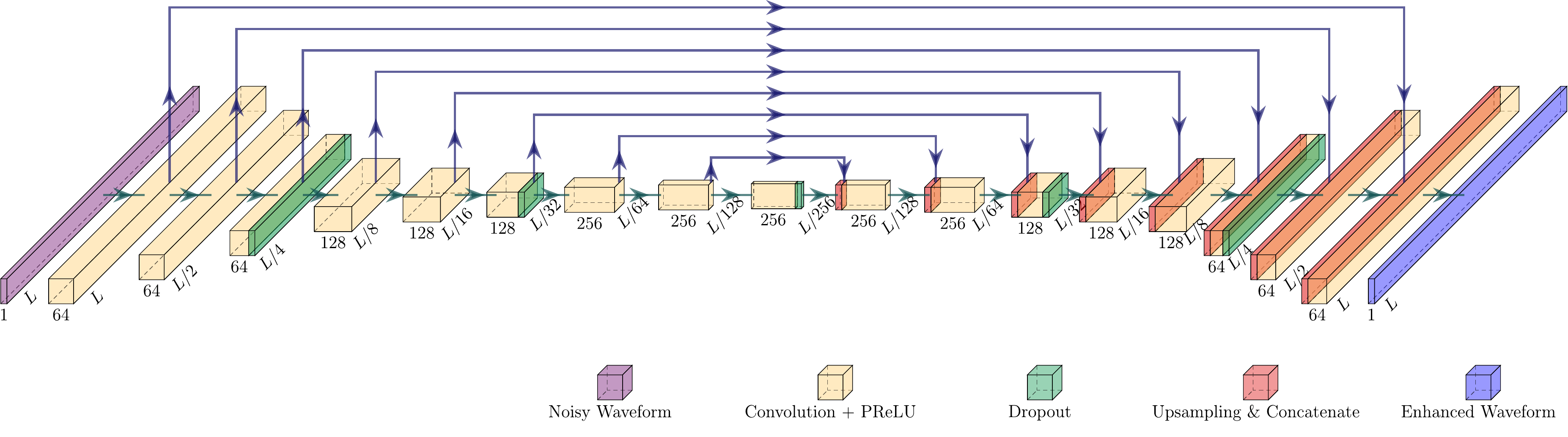}
	\caption{End-to-end speech enhancement system based on a fully convolutional neural network.}
	\label{fig:sefig}
\end{figure*}

In this paper we study the influence of loss functions on the performance of end-to-end time-domain deep learning-based speech enhancement systems.
Specifically, we adopt a general-purpose FCNN architecture that takes as input a time-domain waveform of a noisy speech signal and is trained using various loss functions to predict as output the enhanced speech signal as a time-domain waveform that optimize the loss function in question.

We focus on six loss functions: time-domain mean-square error\,(MSE) $\mathcal{L}_{\text{TIME-MSE}}$, short-time spectral amplitude\,(STSA) MSE $\mathcal{L}_{\text{STSA-MSE}}$ \cite{ephraim_speech_1984}, short-time objective intelligibility\,(STOI) $\mathcal{L}_{\text{STOI}}$ \cite{taal_algorithm_2011}, Extended STOI $\mathcal{L}_{\text{ESTOI}}$ \cite{jensen_algorithm_2016}, scale-invariant signal-to-distortion ratio\,(SI-SDR) $\mathcal{L}_{\text{SI-SDR}}$ \cite{roux_sdr_2019}, and perceptual metric for speech quality evaluation\,(PMSQE) $\mathcal{L}_{\text{PMSQE}}$ \cite{martin-donas_deep_2018}. 
We study these loss functions as they jointly cover a large range of useful properties, e.g. close relationships to human perception or mathematical simplicity, that usually are of interest for speech enhancement systems (described in detail in Sec.\,\ref{sec:sesys}).  
Furthermore, the six loss functions have all been applied in recent deep learning-based speech processing techniques
 e.g. \cite{venkataramani_performance_2018,venkataramani_end--end_2017,fu_end--end_2018,zhao_perceptually_2018,zhang_training_2018,bahmaninezhad_comprehensive_2019,pandey_new_2018,luo_conv-tasnet:_2019,luo_tasnet:_2018-1,kolbaek_monaural_2018-1,kolbaek_relationship_2019,kolbaek_speech_2017,kolbaek_multi-talker_2017-1,wang_supervised_2018,wang_training_2014,erdogan_deep_2017,martin-donas_deep_2018,naithani_deep_2018-1,tan_gated_2019}.
However, no existing work has studied these loss functions jointly under identical conditions and evaluated them in a structured manner with end-to-end time-domain deep learning-based speech enhancement systems.   
Obviously, one would expect, that a system trained to minimize a specific loss function would also achieve the minimum numerical value among \emph{all} systems for that particular loss function during test. 
However, as training of machine learning models in general, and FCNNs in particular, is a highly non-linear process, which depends on the loss function itself, this is not guaranteed in practice. 
Furthermore, we argue that the learning-rate is a crucial design parameter, when conducting such experiments, even for adaptive gradient-based optimizers such as ADAM \cite{kingma_adam:_2015}, which has been generally overlooked in the literature despite its obvious importance.   
Therefore, it is of interest to establish exactly how large the difference in speech enhancement performance is between time-domain waveform-based speech enhancement systems trained using different loss functions and evaluated using popular speech enhancement performance metrics. 
In particular, one might hope to find a "universally good" loss function that performs almost optimally with respect to other loss functions.  
This is the goal of the paper and our findings might serve as a guideline in loss-function selection for deep learning-based speech processing systems. 

The rest of the paper is organized as follows. In Sec.\,\ref{sec:sesys} we describe the monaural speech separation problem and present the six loss functions we will study. In Sec.\,\ref{sec:exdes} we present the design of our experimental study including the speech and noise material used for training. In Sec.\,\ref{sec:exres} we present and discusses the results. Finally, in Sec.\,\ref{sec:con} we conclude the paper.

\section{Speech Enhancement System}\label{sec:sesys}
Fig.\;\ref{fig:sefig} shows a block-diagram of the speech enhancement system we use for all experiments.  
The system is based on a fully convolutional neural network\,(FCNN), which is trained, end-to-end, to estimate the noise-free speech waveform from a noisy single-channel recording. 
Note, the architecture in Fig.\;\ref{fig:sefig} resembles that used in a large body of state-of-the-art deep learning-based speech enhancement literature (e.g. \cite{pandey_new_2018,pandey_tcnn:_2019,baby_sergan:_2019,pascual_segan:_2017,grzywalski_using_2019,tan_real-time_2019,park_fully_2017,ernst_speech_2018}). Therefore, we argue that our experimental findings based on this particular architecture are representative and generally valid for a large range of deep learning-based speech enhancement methods. The architecture in Fig.\;\ref{fig:sefig} is further described in Sec.\,\ref{sec:expmod}.

Let $\barbelow{x} \in \mathbb{R}^L$ be $L$ samples of a clean time-domain speech signal and let the corresponding noisy observation $\barbelow{y} \in \mathbb{R}^L$ be
\begin{equation}
\barbelow{y} = \barbelow{x} + \barbelow{v}, 
\label{eq1}
\end{equation}
where $\barbelow{v} \in \mathbb{R}^L$ is an additive noise signal. The goal is then to find an estimate $\hat{\barbelow{x}}$ of $\barbelow{x}$ from $\barbelow{y}$ using a FCNN,  
\begin{equation}
\hat{\barbelow{x}} = f_{FCNN}(\barbelow{y},\mathbf{\barbelow{\theta}}), 
\label{eq2}
\end{equation}
where $\barbelow{\theta}$ represents the parameters of the FCNN. 
Using a supervised learning approach, the parameters are found such that they minimize a loss $\mathcal{L}$ over a training dataset, consisting of corresponding pairs of clean $\barbelow{x}_{train}$ and noisy speech signals $\barbelow{y}_{train}$.  
Our objective is then to study how the quality of $\hat{\barbelow{x}}$, measured using different performance metrics, is affected by the choice of loss function $\mathcal{L}$.     
In the following, we review each of the loss functions we have selected for our experiments.

\subsection{Time-Domain Mean Square Error}\label{sec:timemse}

The first loss function we consider is the time-domain mean-square error\,(MSE). This loss function is given as 
\begin{equation}
\mathcal{L}_{\text{TIME-MSE}} = \frac{1}{L} \left\lVert \hat{\barbelow{x}} - \barbelow{x} \right\rVert^2_2,
\label{eq3}
\end{equation}
where $\left\lVert \cdot \right\rVert_2$ is the $\ell^2$-norm.
We include this loss function because it is computationally very simple and because it is one of the most used loss functions in machine learning and signal processing in general \cite{loizou_speech_2013,bishop_pattern_2006,goodfellow_deep_2016}. 
However, little is known about the performance of time-domain speech enhancement systems optimized end-to-end for this loss function, when evaluated using standard speech enhancement metrics such as STOI and PESQ.       

\subsection{Short-Time Spectral Amplitude Mean Square Error}\label{sec:freqmse}

The second loss function we consider is the classical STSA-MSE, which is one of the most popular loss functions used for deep neural network based speech enhancement \cite{wang_supervised_2018,kolbaek_relationship_2019}. The STSA-MSE function also plays a major role in more classical non-machine learning based speech enhancement algorithms \cite{ephraim_speech_1984,hendriks_dft-domain_2013}, but has also been used in recent time-domain techniques \cite{pandey_new_2018}.  

Let ${x}(k,m)$,  $k = 1,\dots, K$,  $m=1,\dots M, $ be the $K$-point short-time discrete Fourier transform\,(STFT) of $\barbelow{x}$, where $M = \lfloor \frac{L}{I} \rfloor-1$ is the number of STFT frames with truncation and $I$ is the frame shift in samples.  
Furthermore, let $a(k,m) = |{x}(k,m)| $,  $k = 1,\dots, \frac{K}{2}+1$,  $m=1,\dots M, $ denote the single-sided amplitude spectra of ${x}(k,m)$. Finally, let $\hat{a}(k,m)$ denote the estimate of $a(k,m)$.

The STSA-MSE is then given as
\begin{equation}
\mathcal{L}_{\text{STSA-MSE}} = \frac{1}{\left(\frac{K}{2}+1\right) M} \sum_{k=1}^{\frac{K}{2}+1} \sum_{m=1}^{M} (\hat{a}(k,m) - a(k,m))^2,
\label{eq4}
\end{equation}
which is the MSE between the single-sided amplitude spectra of the true target signal $\barbelow{x}$ and the estimated signal $\hat{\barbelow{x}}$. Note, $\mathcal{L}_{\text{STSA-MSE}}$ is only sensitive to variations in spectral amplitudes and not to variations in the short-time phase spectrum of the signals. 
This is different from $\mathcal{L}_{\text{TIME-MSE}}$ (Eq.\;\eqref{eq3}), as $\mathcal{L}_{\text{TIME-MSE}}$ is operating in the time-domain. 
For all experiments in this paper we use $K=256$ and $I=128$.

\subsection{Short-Time Objective Intelligibility}\label{sec:estoi}

The third loss function we consider is based on the short-time objective intelligibility\,(STOI) speech intelligibility estimator \cite{taal_algorithm_2011}. 
STOI is currently the, perhaps, most commonly used speech intelligibility estimator for objectively evaluating the performance of speech enhancement systems \cite{healy_algorithm_2015,chen_large-scale_2016,healy_algorithm_2017,kolbaek_speech_2017}. 
This is presumably driven by the fact that STOI predictions have shown a good correspondence with measured intelligibility of noisy/processed speech in a large range of acoustic scenarios, including ideal time-frequency weighted noisy speech \cite{taal_algorithm_2011} and noisy speech enhanced by single-microphone time-frequency weighting-based speech enhancement systems \cite{taal_algorithm_2011} (see also \cite{jensen_algorithm_2016,jensen_speech_2014}). 
Therefore, it is natural to believe that gains in speech intelligibility, as estimated by STOI, can be achieved by utilizing a loss function based on STOI.
In the following, we introduce the STOI loss function $\mathcal{L}_{\text{STOI}}$, which essentially is identical to STOI. The main difference is that we omit the voice activity detector\,(VAD) otherwise used by STOI. We do, however, apply the VAD from STOI on the dataset used for training and validation (described further in Sec.\,\ref{sec:exdes}). 

Let $a(k,m)$ $k = 1,\dots, \frac{K}{2}+1$,  $m=1,\dots M, $ denote the single-sided STFT amplitude spectra of the clean speech spectrum as defined in Sec.\,\ref{sec:freqmse}. We then define the $j^{th}$ one-third octave band clean-speech amplitude, for time-frame $m$, as \cite{taal_algorithm_2011}  
\begin{equation}
a_j( m ) = \sqrt{\sum_{k=k_1(j)}^{k_2(j)} a(k,m)^2},
\label{eq5}
\end{equation}
where $k_1(j)$ and $k_2(j)$ denote the first and last STFT bin index, respectively, of the $j^{th}$ one-third octave band.
In a similar fashion we define $\hat{a}_j( m )$ as the $j$th one-third octave band estimated clean-speech amplitude, for time-frame $m$.

Furthermore, let a short-time temporal envelope vector that spans time-frames $m-N+1, \dots, m$, in the $j$th frequency band for the clean speech signal be defined as 
\begin{equation}
\s_{j,m} = [ a_j( m-N+1 ), \; a_j( m-N+2 ), \dots , a_j( m ) ]^T,
\label{eq6}
\end{equation}
where $N=30$, which corresponds to approximately $384$ ms with a sampling frequency of 10 kHz. 

Similarly, we define  $\sh_{j,m}$ as the short-time temporal envelope vector for the enhanced speech signal.
The vector $\sh_{j,m}$ is normalized and clipped for each entry $\sh_{j,m}(n)$ according to 
\begin{equation}
\sh'_{j,m}(n) = \text{min}\left( \frac{ \lVert \s_{j,m} \rVert }{\lVert \sh_{j,m} \rVert }\sh_{j,m}(n), (1+10^{-0.75}) \s_{j,m}(n) \right),
\label{eq7}
\end{equation} 
for $n = 1,2, \dots, N$.

The intermediate intelligibility measure for a pair of short-time temporal envelope vectors $\s_{j,m}$ and $\sh'_{j,m}$ is then defined as the sample linear correlation between the clean and enhanced envelope vectors given as 
\begin{equation}
\sd_{j,m} = \frac{\left(\s_{j,m} - \mu_{\s_{j,m}} \right)^T  \left(\sh'_{j,m} - \mu_{\sh'_{j,m}}\right)}{ \left\lVert \s_{j,m} - \mu_{\s_{j,m}} \right\rVert  \; \left\lVert\sh'_{j,m} - \mu_{\sh'_{j,m}} \right\rVert },
\label{eq8}
\end{equation}
where $\mu_{\s_{j,m}}$ and $\mu_{\sh'_{j,m}}$ are the sample mean vectors of $\s_{j,m}$ and $\sh'_{j,m}$, respectively. 
From $\sd_{j,m}$, the final STOI score for an entire speech signal is then defined as the scalar, $-1 \leq d_{STOI} \leq 1$,    
\begin{equation}
d_{STOI} = \frac{1}{J(M-N+1)} \sum_{j=1}^{J} \sum_{m=N}^{M} \sd_{j,m},
\label{eq9}
\end{equation}
where $J = 15$ is the number of one-third octave bands and $M-N+1$ is the total number of short-time temporal envelope vectors. With $J=15$, the center frequency of the first one-third octave band is 150 Hz and the last one is at approximately 3.8 kHz. These frequencies are chosen such that they span the frequency range in which human speech normally lie\cite{taal_algorithm_2011}.
Finally, with $N=30$, STOI is sensitive to temporal modulation frequencies of $2.6$ Hz and higher, which are frequencies important for speech intelligibility \cite{taal_algorithm_2011}.

We define our STOI loss function to be minimized as
\begin{equation}
\mathcal{L}_{\text{STOI}} = -d_{STOI}.
\label{eq10}
\end{equation}
Note, except for the $\text{min}(\cdot,\cdot)$ operator in Eq.\;\eqref{eq7} the entire STOI loss function is differentiable and computing the required gradients for gradient based optimization is straight forward (see e.g. \cite{kolbaek_relationship_2019}). 
Furthermore, the $\text{min}(\cdot,\cdot)$ operator requires only two subgradients, so the computational complexity of its gradient computation is similar to the standard ReLU activation function, which is nothing more than the max operator. To that end, $\mathcal{L}_{\text{STOI}}$ is suitable as a loss function for training DNN-based speech enhancement systems. %

\subsection{Extended Short-Time Objective Intelligibility}\label{sec:stoi}
The fourth loss function we include is the extended short-time objective intelligibility\,(ESTOI) speech intelligibility estimator \cite{jensen_algorithm_2016}. As the name implies, ESTOI is inspired by STOI and was developed in an attempt to improve STOI. 
Specifically, in \cite{jensen_algorithm_2016} it was shown that the performance of certain speech intelligibility estimators, including STOI, was sensitive to spectro-temporal modulations of the noise component and that STOI did not correlate as well ($\rho = 0.47$ \cite{jensen_algorithm_2016}) with listening test results, when the noise components were highly fluctuating (as e.g. with a competing talker). 

To alleviate this drawback of STOI, ESTOI was proposed \cite{jensen_algorithm_2016}. It was shown that ESTOI significantly outperformed STOI ($\rho > 0.90$ \cite{jensen_algorithm_2016}), as well as other speech intelligibility estimators, in conditions when the noise type is highly fluctuating, while performing on par with these estimators in less fluctuating noise conditions. 
Consequently, it is of interest to study how ESTOI compares with STOI, as a loss function, for time-domain DNN-based speech enhancement.  

Similarly to STOI, ESTOI is based on an average correlation coefficient between one-third octave band short-time temporal envelope vectors.         
Specifically, let,
\begin{gather}
\barbelow{\barbelow{A}}_m
=
\begin{bmatrix}
a_1( m-N+1 ) & \dots & a_1(m) \\
\vdots       &       & \vdots \\
a_{J}( m-N+1 ) & \dots & a_{J}(m) \\
\end{bmatrix}
\label{eq11}
\end{gather}
denote a short-time spectrogram matrix of the clean speech signal, where the rows of $\barbelow{\barbelow{A}}_m$ are given by $\s_{j,m}$, which are short-time temporal envelope vectors in a one-third band defined by Eq.\,\eqref{eq6}.
The $j$th mean- and variance-normalized row of $\barbelow{\barbelow{A}}_m$ is then given by    
\begin{equation}
\bar{\s}_{j,m} = \frac{1}{\lVert  (\s_{j,m}  -  \mu_{\s_{j,m}} )    \lVert} (\s_{j,m}  -  \mu_{\s_{j,m}} ).
\label{eq12}
\end{equation}
%
ESTOI now introduces the row-normalized spectrogram matrix 
\begin{gather}
\barbelow{\barbelow{\bar{A}}}_m
=
\begin{bmatrix}
\bar{\s}^T_{1,m} \\
\vdots        \\
\bar{\s}^T_{J,m}\\
\end{bmatrix}_,
\label{eq13}
\end{gather}

and defines $\check{\s}_{n,m}$ as the mean- and variance-normalized $n$th column, $n=1,2,\dots,N$ of $\barbelow{\barbelow{\bar{A}}}_m$, where the normalization of the columns is performed analogously to Eq.\,\eqref{eq12}.

Finally, define
\begin{gather}
\barbelow{\barbelow{\check{A}}}_{m}
=
\left[\check{\s}_{1,m} \dots \check{\s}_{N,m}  \right]
\label{eq14}
\end{gather}
as the row and column normalized spectrogram matrix. Similarly, we define $\hat{\s}^{''}_{n,m}$ as the columns of the row and column normalized spectrogram matrix for the enhanced speech signal $\barbelow{\barbelow{\hat{A}}}_{m}$.
Finally, the ESTOI speech intelligibility index is defined as
\begin{equation}
d_{ESTOI} = \frac{1}{NM} \sum_{m=1}^{M} \sum_{n=1}^{N} \check{\s}^T_{n,m}  \hat{\s}^{''}_{n,m}.
\label{eq15}
\end{equation}
Similarly to $\mathcal{L}_{\text{STOI}}$, we define the ESTOI loss as
\begin{equation}
\mathcal{L}_{\text{ESTOI}} = -d_{ESTOI}.
\label{eq16}
\end{equation}
Note, differently from $\mathcal{L}_{\text{STOI}}$, $\mathcal{L}_{\text{ESTOI}}$ does not include the clipping step, i.e. the  $\text{min}(\cdot,\cdot)$ operator in Eq.\;\eqref{eq7}, which makes $\mathcal{L}_{\text{ESTOI}}$ fully differentiable. Also, similarly to the definition of $\mathcal{L}_{\text{STOI}}$, we have ignored the VAD otherwise used by $\mathcal{L}_{\text{ESTOI}}$ as we apply the VAD on the data prior to training.

\subsection{Scale-Invariant Signal-to-Distortion Ratio}\label{sec:sisdr}
The fifth loss function we include is the scale-invariant signal-to-distortion ratio\,(SI-SDR) \cite{roux_sdr_2019}.    
The SI-SDR is an objective performance measure that was introduced for evaluating the performance of speech processing algorithms and it was proposed as an alternative to the often used SDR measure from the \emph{BSS\_eval} toolbox \cite{fevotte_bss_2011}. 
Differently from SDR, SI-SDR is invariant to the scale of the processed signal, but not to deformations caused by finite-impulse response filters as SDR is \cite{roux_sdr_2019}.     

The SI-SDR is defined as
\begin{equation}
\begin{split}
\text{SI-SDR} & = 10 \; \text{log}_{10} \left( \frac{\lVert \alpha \barbelow{x} \lVert^2 }{\lVert \alpha \barbelow{x} - \hat{\barbelow{x}}  \lVert^2} \right),
\end{split}
\label{eq17_1}
\end{equation}
where  
\begin{equation}
\alpha = \frac{\hat{\barbelow{x}}^T \barbelow{x} }{\lVert  \barbelow{x} \lVert^2}   = \underset{\alpha}{\text{argmin}}  \lVert \alpha \barbelow{x} - \hat{\barbelow{x}}  \lVert^2.
\label{eq18}
\end{equation}
It is seen from Eqs.\,\eqref{eq17_1}, that SI-SDR is simply the signal-to-noise\,(SNR) ratio between the weighted clean speech signal and the residual noise defined as $\lVert \alpha \barbelow{x} - \hat{\barbelow{x}}  \lVert^2$. 
Hence,
\begin{equation}
\begin{split}
\text{SI-SDR} & = 10 \; \text{log}_{10} \left( \frac{ \left\lVert \frac{\hat{\barbelow{x}}^T \barbelow{x} }{\lVert  \barbelow{x} \lVert^2}  \barbelow{x} \right\lVert^2 }{ \left\lVert \frac{\hat{\barbelow{x}}^T \barbelow{x} }{\lVert  \barbelow{x} \lVert^2}  \barbelow{x} - \hat{\barbelow{x}}  \right\lVert^2 }  \right)\\
& = 10 \; \text{log}_{10} \left( \frac{\barbelow{x}^T\hat{\barbelow{x}}}{\barbelow{x}^T\barbelow{x}\hat{\barbelow{x}}^T\hat{\barbelow{x}} -\barbelow{x}^T\hat{\barbelow{x}}} \right).
\end{split}
\label{eq17_2}
\end{equation}
The scaling of the reference signal $\barbelow{x}$ ensures that the SI-SDR measure is invariant to the scale of $\hat{\barbelow{x}}$, which might be desirable in applications, where the speech processing algorithm do not guarantee a proper scaling of the processed signal, such as many DNN-based systems. This is also motivated by the fact that both speech quality and intelligibility to a large extent is invariant to scaling \cite{moore_introduction_2013}.

Note, that maximizing $\mathcal{L}_{\text{SI-SDR}}$ is equivalent to maximizing the sample correlation between $\barbelow{x}$ and $\hat{\barbelow{x}}$, while producing the solution with the minimum energy \cite{venkataramani_performance_2018,venkataramani_end--end_2017}.    
Furthermore, similarly to SNR, SI-SDR is expressed in units of decibel\,(dB) and  is defined in the range $-\infty < \text{SI-SDR} < \infty$, which motivates us to define the SI-SDR loss function as  
\begin{equation}
\mathcal{L}_{\text{SI-SDR}} = -\text{SI-SDR}.
\label{eq19}
\end{equation}

\subsection{Perceptual Metric for Speech Quality Evaluation}\label{sec:pmsqe}
The sixth, and last loss function is the perceptual metric for speech quality evaluation\,(PMSQE) \cite{martin-donas_deep_2018}. 
The PMSQE loss function, $\mathcal{L}_{\text{PMSQE}}$, is designed to approximate the non-differentiable perceptual evaluation of speech quality\,(PESQ) speech quality estimator. The PESQ speech quality estimator is furthermore designed to predict the mean opinion score\,(MOS) of a speech quality listening test for certain degradations. Consequently, the PESQ score of a processed speech signal is a scalar between $1$ and $4.5$, where $1$ indicates extremely poor quality and $4.5$ corresponds to no distortion at all \cite{rix_perceptual_2001,noauthor_international_2003}. 
Along the same lines, the PMSQE loss function is designed to be inversely proportional to PESQ, such that a low PMSQE value corresponds to a high PESQ value and vice versa. 
In practice PMSQE is defined in the range from $3$ to $0$, where $0$ is equivalent to an undistorted signal and $3.0$ corresponds to an extremely poor quality. 

\begin{figure}
	\centering
	\includegraphics[trim={8mm 0mm 4mm 0mm},clip,width=1.0\linewidth]{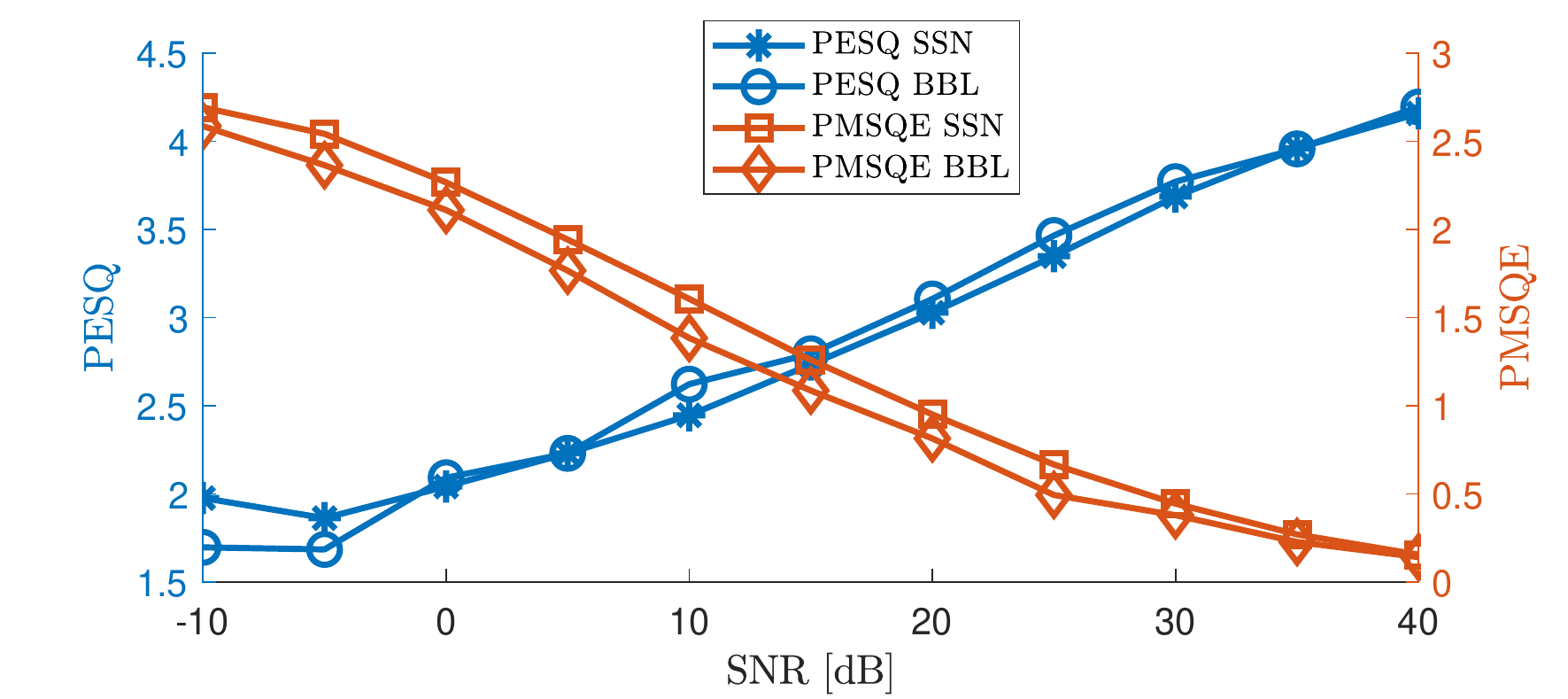}
	\caption{PESQ ITU P.862.1 and PMSQE scores as function of SNR for SSN and BBL noise-corrupted speech. }
	\label{fig:pesq_vs_pmsqe}
\end{figure}

Fig.\,\ref{fig:pesq_vs_pmsqe} shows the correspondence between PESQ and PMSQE for a speech signal corrupted with either a stationary speech shaped noise\,(SSN) signal or a non-stationary 6-speaker babble noise signal, at various SNRs. 
It is seen that PESQ and PMSQE are approximately inversely proportional and have a monotonic relationship with respect to SNR. Hence, it is assumed that if PMSQE is minimized, PESQ will be maximized. 
The $\mathcal{L}_{\text{PMSQE}}$ loss function is essentially a log-domain STSA-MSE loss function with additional key terms that are inspired by human perception. Consequently, an outline of $\mathcal{L}_{\text{PMSQE}}$ is rather involved, and we refer the reader to \cite{martin-donas_deep_2018} for details regarding the design of PMSQE. 
Furthermore, as PMSQE, similarly to PESQ, is defined for sampling rates at either 8 kHz or 16 kHz, we use a 8 kHz sampling frequency when training $\mathcal{L}_{\text{PMSQE}}$ systems, and we downsample test signals to 8 kHz, when we evaluate speech enhancement systems using PESQ.            

\section{Experimental Design}\label{sec:exdes}
To study how the loss functions presented in Sec.\,\ref{sec:sesys} affect the performance of FCNN-based speech enhancement systems in realistic acoustical conditions, we train multiple systems using a large noisy-speech dataset with a high degree of speaker and noise variability. In the following, we introduce the dataset, noise types and mixture conditions used for all experiments presented in Sec.\,\ref{sec:exres}.

\subsection{Noise-free Speech Mixtures}
We have evaluated the six loss functions using the WSJ0 speech corpus \cite{garofolo_csr-i_1993}. Specifically, using a sampling-with-replacement scheme, the training data is based on 30000 randomly selected spoken utterances from a subset of the \emph{si\_tr\_s} part of WSJ0. 
The dataset size was found during preliminary experiments to be a good trade-off between training time and speech enhancement performance. 
This \emph{si\_tr\_s} subset of WSJ0 consists in total of 11613 utterances approximately equally divided among 44 male speakers and 47 female speakers. This ensures that the training dataset contains a large speaker variability, which allows the final speech enhancement system to be largely speaker independent \cite{kolbaek_speech_2017}. 

Similarly, the validation set is based on 3000 randomly selected spoken utterances from another subset of \emph{si\_tr\_s}, which consists of 1163 spoken utterances divided among five male speakers and five female speakers, which are not present in the training set. 

Finally, the test set is based on 1000 randomly selected spoken utterances from \emph{si\_et\_05} and \emph{si\_dt\_05}, which consists of 1857 utterances divided among ten males and six females.     
  
Note, as the training and validation sets consist of approximately three times as many utterances as their respective subsets of \emph{si\_tr\_s}, each utterance from WSJ0 will on average be selected three times. However, as each utterance is mixed with its own unique noise signal, the redundancy in speech material  increases the total variability in the dataset and ultimately improves the generalizability capability of the system. 
Also note that the speakers used in the training and validation sets are different than the speakers used for test, i.e. the tests are conducted in a speaker independent setting.  

Furthermore, as we are primarily interested in speech active regions during training, we apply the voice activity detector\,(VAD) from STOI (and ESTOI) \cite{taal_algorithm_2011,jensen_algorithm_2016} on the training and validation set to ensure that any potentially long silent regions are removed prior to training. 
Specifically, the VAD  analyzes the clean waveform in 25 ms segments and removes the segments where the signal energy is more than 40 dB below the energy of the segment with the maximum energy in the waveform.

Finally, all utterances used with $\mathcal{L}_{\text{STOI}}$, $\mathcal{L}_{\text{ESTOI}}$, $\mathcal{L}_{\text{TIME-MSE}}$ and $\mathcal{L}_{\text{SI-SDR}}$ are downsampled to 10 kHz, as STOI and ESTOI are defined for this sampling frequency, and to allow an efficient training scheme using minibatch training, each utterance is truncated or zero-padded\footnote{zero-padding constitutes only $3.9\;\%$ of the total number of samples.} to four seconds. The utterances used with $\mathcal{L}_{\text{PMSQE}}$ are downsampled to 8 kHz to comply with the definition of PMSQE, which results in an utterance duration of approximately five seconds.

\subsection{Noise Types}
To ensure a diverse noise variability we include four different noise types in the training dataset: two synthetic noise signals and two real-life recordings of natural sound scenes. 
This is motivated by the fact that \emph{a priori} knowledge about the noise type might lead to unrealistic performance estimates \cite{kolbaek_speech_2017}.  
The two synthetic noise signals are a stationary speech shaped noise\;(SSN) and a non-stationary 6-speaker babble\;(BBL) noise. 
The SSN signal is synthetically generated Gaussian white noise that is spectrally shaped using a $12$th-order all-pole filter with coefficients found from linear predictive coding analysis of the concatenation of 100 randomly chosen TIMIT sentences \cite{garofolo_timit_1993}.  
The BBL noise signal is constructed as a linear mix of randomly selected utterances from the TIMIT corpus such that six speakers are speaking at any given time. Using the entire TIMIT database of 6300 utterances results in a BBL noise sequence with a duration of more than 50 min. 
For the real-life noise signals, we use the street\;(STR) and cafeteria\;(CAF), noise signals from the CHiME3 dataset, which are signals that have been recorded in a natural occurring sound scene \cite{barker_third_2015}.    

Finally, we divide the noise signals such that 40 minutes is used for training, five minutes is used for validation and another five minutes is used for test. This ensures that each noise type is equally represented and with unique realizations in each dataset.    

To evaluate the performance of the speech enhancement systems to unseen or unmatched noise signals we also test using the bus\;(BUS), and pedestrian\;(PED) noise signals from \cite{barker_third_2015}. These noise signals are also real-life recordings, but they represent different noise statistics compared to the four noise types used for training.     

\subsection{Noisy Speech Mixtures}
To construct the noisy speech signals, we follow Eq.\;\eqref{eq1} and combine a noise-free training utterance $\barbelow{x}$ with an equal length and randomly selected noise sequence $\barbelow{v}$. 
The noise signal $\barbelow{v}$ is scaled according to the active speech level of $\barbelow{x}$ as defined by ITU P.56 \cite{itu_rec._2011} to achieve a certain SNR.
For the training and validation datasets, this SNR is chosen uniformly from $[-10 , 10 ]$, which ensures that the intelligibility of the noisy speech waveforms $\barbelow{y}$ ranges from poor to perfectly intelligible.

\subsection{Model Architecture and Training} \label{sec:expmod}
The speech enhancement system (Fig.\;\ref{fig:sefig}) consists of a FCNN with 18 layers configured in an encoder/decoder architecture \cite{ronneberger_u-net:_2015} using parameterized ReLU (PReLU) activation functions \cite{he_delving_2015}. 
The input dimension is $L=38656$ and except for the first layer all remaining layers in the encoder use a stride of two, which drives the final dimension in the bottleneck to be of dimension $L/256$. Similarly, except for the last layer, which has dimension $L$, all layers in the decoder uses upsampling with a factor of two. 
Additionally, skip-connections where incoming channels are concatenated with existing channels are used between the first eight layers in the encoder and the corresponding eight layers in the decoder. Similarly to \cite{pandey_new_2018}, during training 20\;\% dropout is used for every third layer.

Furthermore, in (inChannel, outChannel, stride) format, the FCNN model has one (1,64,1), two (64,64,2),  one (64,128,2), two (128,128,2), one (128,256,2), two (256,256,2), two (512,256,1), three (256,128,1), three (128,64,1), and one (128,1,1) convolutional layers with a filter size of 11 samples, which makes the model comparable to other enhancement models in the literature (see e.g. \cite{pascual_segan:_2017,pandey_new_2018,pandey_new_2019}). In total, the model has approximately 6.8 million parameters.

Note, due to the encoder/decoder architecture, the receptive field is 2561 samples, which means that 2561 samples need to be available before the system can produce a single output. In other words, with a 10 kHz sampling frequency the latency of the speech enhancement system is 256 ms. For applications where hard real-time requirements apply, e.g. hearing aids, this latency can likely be reduced significantly using alternative architectures (e.g. \cite{luo_tasnet:_2018-1}).    

The speech enhancement system  is trained using the ADAM optimizer \cite{kingma_adam:_2015} with $\beta_1 = 0.9$ and $\beta_2=0.999$ and a learning rate schedule that reduces the learning rate with a factor of two, if the validation loss has not decreased for two epochs. 
The six loss functions considered in this study have different gradients and ultimately different gradient norms. Consequently, a learning rate used for one loss function might not be the optimal learning rate for another loss function. In fact, using a non-optimal learning rate might result in radical different solutions and potentially erroneous conclusions, as we show in Sec.\,\ref{sec:exresLR}. 
We use optimized (and different) learning rates for the different loss functions as further described in Sec.\,\ref{sec:exresLR}. The learning rates are shown in Table\;\ref{tab:models}. Finally, a batch size of eight is used, and training is stopped, if the validation loss has not decreased for five epochs or a maximum of 200 epochs has elapsed.

We have implemented the speech enhancement systems using Keras\footnote{https://keras.io/} with a TensorFlow\footnote{https://tensorflow.org/} backend and the python implementation of the models and loss functions, as well as audio samples, are available online\footnote{\url{https://git.its.aau.dk/mok/Speech_Enhancement_Loss.git}}.

\section{Experimental Results}\label{sec:exres}
We now investigate empirically how each of the loss functions presented in Sec.\,\ref{sec:sesys} affects the speech enhancement performance of the time-domain FCNN-based speech enhancement system presented in Sec.\,\ref{sec:exdes}. 
Specifically, in Sec.\,\ref{sec:exresLR} we study the sensitivity of speech enhancement performance with respect to learning rate. Such a study is a prerequisite to allow a fair comparison between the custom loss functions in subsequent studies. We then study in Sec.\,\ref{sec:exresSI} how the signal integrity varies among the loss functions. 
Lastly, in Sec.\,\ref{sec:exresLS}, we study the speech enhancement performance for each loss function in various both matched and unmatched noise types at a wide range of SNRs.   
We evaluate the speech enhancement performance of all the systems using the following popular and often used metrics: STOI \cite{taal_algorithm_2011}, ESTOI \cite{jensen_algorithm_2016}, SI-SDR \cite{roux_sdr_2019}, SDR \cite{fevotte_bss_2011}, and PESQ \cite{rix_perceptual_2001}.

\subsection{Learning Rate vs. Performance Metric}\label{sec:exresLR}
Since the goal of this paper is to make a comparison between loss functions, it is important that the comparison is just. However, as the loss functions presented in Sec.\,\ref{sec:sesys} have different processing steps, they have different partial derivatives, which might lead to different gradient norms and a varying sensitivity to the choice of learning rate during gradient based optimization (e.g. \cite{kolbaek_relationship_2019,liu_variance_2020}).       
Therefore, to study the influence that the learning rate can have on the performance of time-domain FCNN-based speech enhancement systems, we have trained multiple systems with various learning rates. Specifically, for each of the six loss functions in Sec.\,\ref{sec:sesys}, we have trained a system using the following five learning rates: $10^{-2}$, $10^{-3}$, $5\cdot10^{-4}$, $10^{-4}$, and $10^{-5}$. 
The learning rates have been selected from preliminary experiments in order to cover the two training extremes, when training either diverge, i.e. a too large learning rate is used, or when training converge too slowly and ultimately ends up at a plateau with a validation loss higher than the validation loss achieved using a larger learning rate. 
The systems for this particular experiment have been trained using SSN at and SNR of 0 dB\footnote{Preliminary experiments using BBL indicated similar results.}.           

In Table\;\ref{tab:lr_res} we present different performance scores for time-domain FCNN-based speech enhancement systems trained using different loss functions and learning rates. The largest performance scores with respect to each loss function (i.e. column-wise) is highlighted in boldface. 
It is eminent from Table\;\ref{tab:lr_res} that a learning rate of $10^{-2}$ is too large for all loss functions as none of the loss functions manage to improve the validation loss. 
Similarly, it is seen from Table\;\ref{tab:lr_res} that a learning rate of $10^{-5}$ is too small for all loss functions as none of the systems, except for $\mathcal{L}_{\text{SI-SDR}}$ evaluated using STOI, achieve the largest scores for this particular learning rate. However, the $\mathcal{L}_{\text{SI-SDR}}$ systems achieve the same STOI score for the three smallest learning rates.
Furthermore, it is seen that the learning rates in the middle range, i.e. $5\cdot10^{-4}$, and $10^{-4}$ achieve particularly large scores. 
Specifically, it is seen from Table\;\ref{tab:lr_res} that $\mathcal{L}_{\text{TIME-MSE}}$, $\mathcal{L}_{\text{SI-SDR}}$, $\mathcal{L}_{\text{STSA-MSE}}$, and $\mathcal{L}_{\text{PMSQE}}$ all achieve the largest overall performance scores using a learning rate of $5\cdot10^{-4}$, whereas the remaining loss functions $\mathcal{L}_{\text{STOI}}$ and $\mathcal{L}_{\text{ESTOI}}$ achieve their maximum performance scores with a learning rate of $10^{-4}$.

More importantly, it is seen that choosing a non-optimal learning rate might actually lead to a wrong conclusion, if the systems were compared based on the same learning rate. This is a consideration that has been generally absent in the literature.  
For example, with the standard\footnote{value originally proposed in \cite{kingma_adam:_2015} and currently default in \url{https://keras.io/}.} learning rate of $10^{-3}$ the $\mathcal{L}_{\text{TIME-MSE}}$ and $\mathcal{L}_{\text{ESTOI}}$ systems both achieve an ESTOI score of 0.79, which might lead to the, perhaps faulty, conclusion that both loss functions possess the same potential with respect to ESTOI improvements.   
However, with a learning rate of e.g. $5\cdot10^{-4}$, it is seen that the $\mathcal{L}_{\text{TIME-MSE}}$ system still achieves an ESTOI score of 0.79, whereas the $\mathcal{L}_{\text{ESTOI}}$ system achieves a considerably larger ESTOI score of 0.83, which leads to the correct conclusion that the $\mathcal{L}_{\text{ESTOI}}$ loss function has potential to outperform the $\mathcal{L}_{\text{TIME-MSE}}$ in terms of ESTOI. 
Similar observations can be made for other loss functions e.g. with respect to $\mathcal{L}_{\text{TIME-MSE}}$ and $\mathcal{L}_{\text{SI-SDR}}$. 
Furthermore, it is seen that the SI-SDR scores have a large variance and with a learning rate of e.g. $10^{-5}$ they can vary from $-22.36$ dB for systems optimized for $\mathcal{L}_{\text{ESTOI}}$ to $10.15$ dB for systems optimized for $\mathcal{L}_{\text{SI-SDR}}$, while achieving comparable STOI, ESTOI, and PESQ scores. This phenomenon is somewhat surprising and is further studied in Sec.\,\ref{sec:exresSI}.
Also, Table\;\ref{tab:lr_res} suggests that when a system is trained with a specific loss function, no other system achieves a larger performance score with respect to that particular metric. This expected result indicates that training has evolved correctly and that the learning rates used in Table\;\ref{tab:lr_res} are close to optimal.     
Finally, it is seen that although $\mathcal{L}_{\text{PMSQE}}$ is an approximation of PESQ, in Table\;\ref{tab:lr_res}, $\mathcal{L}_{\text{SI-SDR}}$ and $\mathcal{L}_{\text{STSA-MSE}}$ consistently lead to larger PESQ scores than $\mathcal{L}_{\text{PMSQE}}$ despite $\mathcal{L}_{\text{PMSQE}}$ consistently achieving the lowest PMSQE values of the two loss functions. $\mathcal{L}_{\text{PMSQE}}$ does, however, lead to larger PESQ scores than $\mathcal{L}_{\text{STOI}}$ and $\mathcal{L}_{\text{ESTOI}}$ for several testing conditions.

In conclusion, selecting the learning rate can have a profound impact on the performance of FCNN-based speech enhancement systems and selecting the proper learning rate is crucial, when systems trained using different loss functions are compared. 
Table\;\ref{tab:models} summarizes the learning rates that we will use for training the systems presented in Sec.\;\ref{sec:exresLS}. The learning rates are selected as the ones that maximize the performance metric most similar to the loss function the systems were trained with. 
Please note that we used four significant digits when the learning rates were selected to ensure a proper resolution.

\begin{table}
	\caption{Performance of different speech enhancement systems measured using various performance metrics. The systems have been tested in matched noise-type conditions using SSN at 0 dB SNR.}	
	\label{tab:lr_res}
	\centering
	\setlength\tabcolsep{4pt} 
	\resizebox{1.00\columnwidth}{!}{%
		\begin{tabular}{cllcccccc}
			\midrule
			\multirow{2}{*}{\begin{tabular}[]{@{}c@{}} Learning \\ Rate \end{tabular} } & \multirow{2}{*}{Metric} & \multirow{2}{*}{Noisy} &\multicolumn{6}{c}{Processed}  \\ \cmidrule(lr){4-9}
			&  &  &$\mathcal{L}_{\text{TIME-MSE}}$ & $\mathcal{L}_{\text{STOI}}$ & $\mathcal{L}_{\text{ESTOI}}$ & $\mathcal{L}_{\text{SI-SDR}}$ & $\mathcal{L}_{\text{STSA-MSE}}$ & $\mathcal{L}_{\text{PMSQE}}$ \\	
			\midrule 
			\multirow{ 6}{*}{$10^{-2}$} 
			& STOI:  & 0.75     & \multicolumn{6}{c}{\multirow{6}{*}{ \begin{tabular}[]{@{}c@{}} Could not improve training or validation loss \\ (No convergence) \end{tabular} }  }  \\ 
			& ESTOI: & 0.46     &  \\ 
			& SI-SDR:& -1.05    & \\ 
			& SDR:   & -0.92    & \\ 
			& PMSQE: & 2.53		& \\
			& PESQ:  & 1.79		& \\
			\midrule 
			\multirow{ 6}{*}{$10^{-3}$} 
			& STOI:  & 0.75     & \textbf{0.92} 	& \textbf{0.93} 	& 0.91 				& 0.91 				& \textbf{0.92} 	& \textbf{0.89}\\ 
			& ESTOI: & 0.46     & \textbf{0.79} 	& 0.81 				& 0.79 				& 0.77 				& 0.79 				& 0.73\\ 
			& SI-SDR:& -1.05    & 10.24 			& \textbf{3.55}		& -4.37				& 9.82 				& 6.59 				& \textbf{-1.07}\\ 
			& SDR:   & -0.92    & 10.88				& 6.47  			& 4.77  			& 10.52 			& 9.39  			& 1.89\\ 
			& PMSQE: & 2.53		& 1.10				& 1.47				& 1.48				& 1.20				& 1.22				& 1.14\\
			& PESQ:	 & 1.79	    & 2.72 				& 2.67 				& 2.51				& 2.65 				& 2.77 				& 2.65\\
			\midrule 
			\multirow{ 6}{*}{$5\cdot10^{-4}$} 
			& STOI:  & 0.75     & \textbf{0.92} 	& \textbf{0.93} 	& \textbf{0.93} 	& \textbf{0.92} 	& \textbf{0.92} 	& \textbf{0.89}\\ 
			& ESTOI: & 0.46     & \textbf{0.79} 	& \textbf{0.82} 	& \textbf{0.83} 	& \textbf{0.80} 	& \textbf{0.80} 	& \textbf{0.75}\\ 
			& SI-SDR:& -1.05    & \textbf{10.30} 	& 2.09 				& \textbf{3.12}		& \textbf{10.70}	& -4.32 			& -8.52\\ 
			& SDR:   & -0.92    & \textbf{11.00}	& \textbf{7.73}		& \textbf{7.84}		& \textbf{11.32} 	& 2.27  			& 4.21\\ 
			& PMSQE: & 2.53 	& 1.07				& 1.43				& 1.27				& 1.01				& 1.19				& \textbf{1.05}\\
			& PESQ:	 & 1.79  	&		  2.73 		& 2.70 				& 2.68 				& \textbf{2.77}		& \textbf{2.80}		& \textbf{2.72}\\
			\midrule 
			\multirow{ 6}{*}{$10^{-4}$} 
			& STOI:  & 0.75     & \textbf{0.92} 	& \textbf{0.93} 	& \textbf{0.93} 	& \textbf{0.92} 	& \textbf{0.92} 	& \textbf{0.89}\\ 
			& ESTOI: & 0.46     & \textbf{0.79} 	& \textbf{0.82} 	& \textbf{0.83} 	& \textbf{0.80} 	& \textbf{0.80} 	& 0.74\\ 
			& SI-SDR:& -1.05    & 10.13 			& 1.95 				& -12.03			& 10.59				& \textbf{8.10} 	& -6.96\\ 
			& SDR:   & -0.92    & 10.78				& 4.99  			& 0.61  			& 11.22 			& \textbf{9.46}  	& \textbf{4.76}\\ 
			& PMSQE: & 2.53		& 1.11				& 1.46				& 1.24				& 1.03				& 1.21				& 1.12\\
			& PESQ:	 & 1.79		& 2.72 				& 2.69 				& 2.68 				& \textbf{2.77}		& 2.79 				& 2.67\\
			\midrule 
			\multirow{ 6}{*}{$10^{-5}$} 
			& STOI:  & 0.75     & 0.90 				& 0.92 				& 0.92 				& \textbf{0.92} 	& 0.91 				& 0.85\\ 
			& ESTOI: & 0.46     & 0.74 				& 0.80 				& 0.81 				& 0.79 				& 0.77 				& 0.67\\ 
			& SI-SDR:& -1.05    & 8.78  			&-2.22  			& -22.36			& 10.15 			& -3.51 			& -6.52\\ 
			& SDR:   & -0.92    & 9.46				& 7.28  			& 4.81  			& 10.78 			& 4.76  			& -0.65\\ 
			& PMSQE: & 2.53		& 1.40				& 1.56				& 1.47				& 1.12				& 1.36			 	& 1.51\\
			& PESQ:	 & 1.79 	& 2.53				& 2.58				& 2.59 				& 2.72 				& 2.69 				& 2.37\\
			\midrule 
	\end{tabular}}
\end{table}

\begin{table}
	\caption{Optimal learning rates for different loss functions.}
	\label{tab:models}
	\centering
	\resizebox{1.0\columnwidth}{!}{
		\begin{tabular}{l|cccccc}
			\toprule
			Loss:  	& $\mathcal{L}_{\text{TIME-MSE}}$ & $\mathcal{L}_{\text{STOI}}$ & $\mathcal{L}_{\text{ESTOI}}$ & $\mathcal{L}_{\text{SI-SDR}}$ & $\mathcal{L}_{\text{STSA-MSE}}$ & $\mathcal{L}_{\text{PMSQE}}$ \\ 
			LR:		& $5\cdot10^{-4}$ & $10^{-4}$ & $10^{-4}$ & $5\cdot10^{-4}$ & $5\cdot10^{-4}$ & $5\cdot10^{-4}$ \\\midrule
	\end{tabular}}
\end{table}

\begin{figure}
	\centering
	\begin{minipage}{.49\columnwidth}
		\centering
		\centerline{\includegraphics[trim={0mm 0mm 0mm 0mm},clip,width=0.95\linewidth]{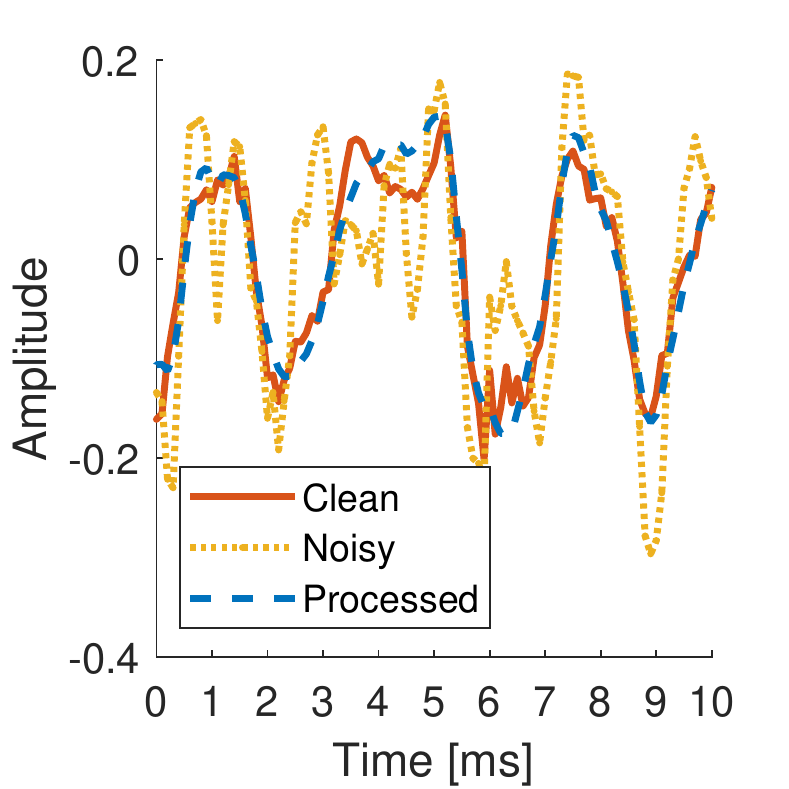}}
		\subcaption{$\mathcal{L}_{\text{TIME-MSE}}$}
		\label{fig:fig1104}
	\end{minipage}%
	\begin{minipage}{.49\columnwidth}
		\centering
		\centerline{\includegraphics[trim={0mm 0mm 0mm 0mm},clip,width=0.95\linewidth]{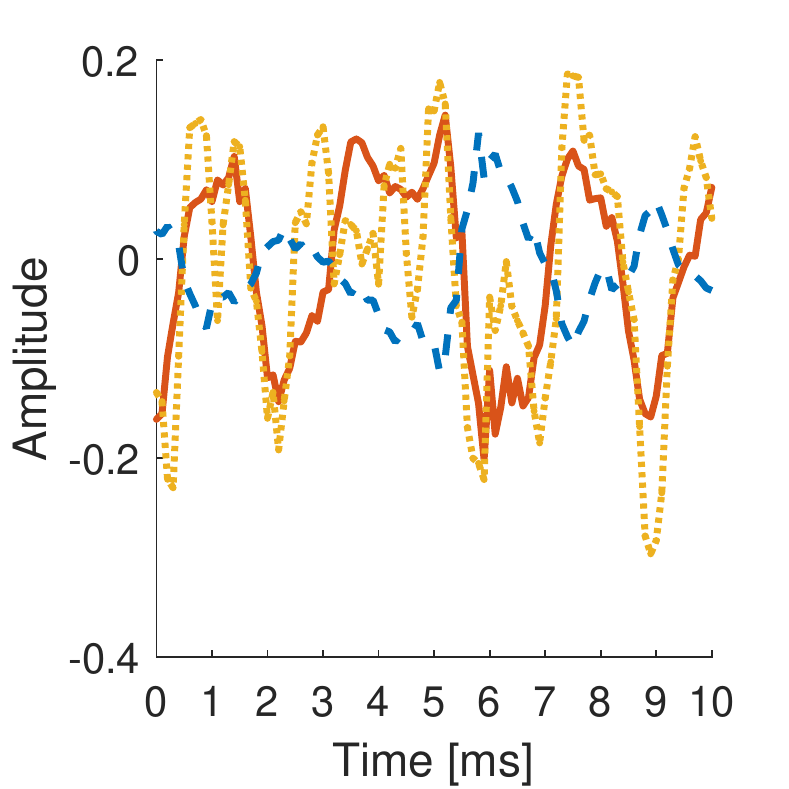}}
		\subcaption{$\mathcal{L}_{\text{STOI}}$}
		\label{fig:fig2104}
	\end{minipage}
	\begin{minipage}{.49\columnwidth}
		\centering
		\centerline{\includegraphics[trim={0mm 0mm 0mm 0mm},clip,width=0.95\linewidth]{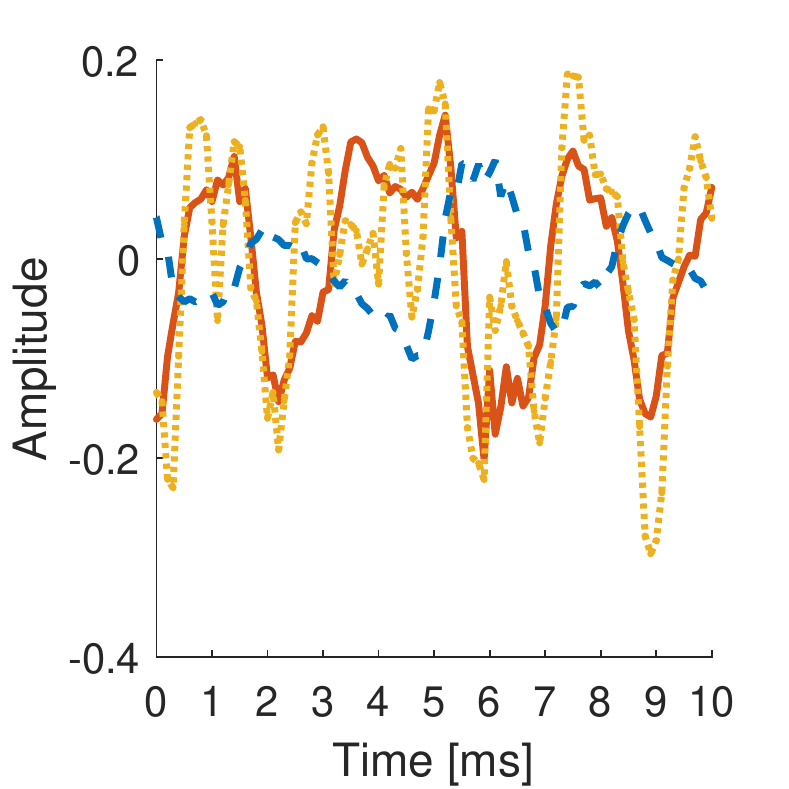}}
		\subcaption{$\mathcal{L}_{\text{ESTOI}}$}
		\label{fig:fig3104}
	\end{minipage}%
	\begin{minipage}{.49\columnwidth}
		\centering
		\centerline{\includegraphics[trim={0mm 0mm 0mm 0mm},clip,width=0.95\linewidth]{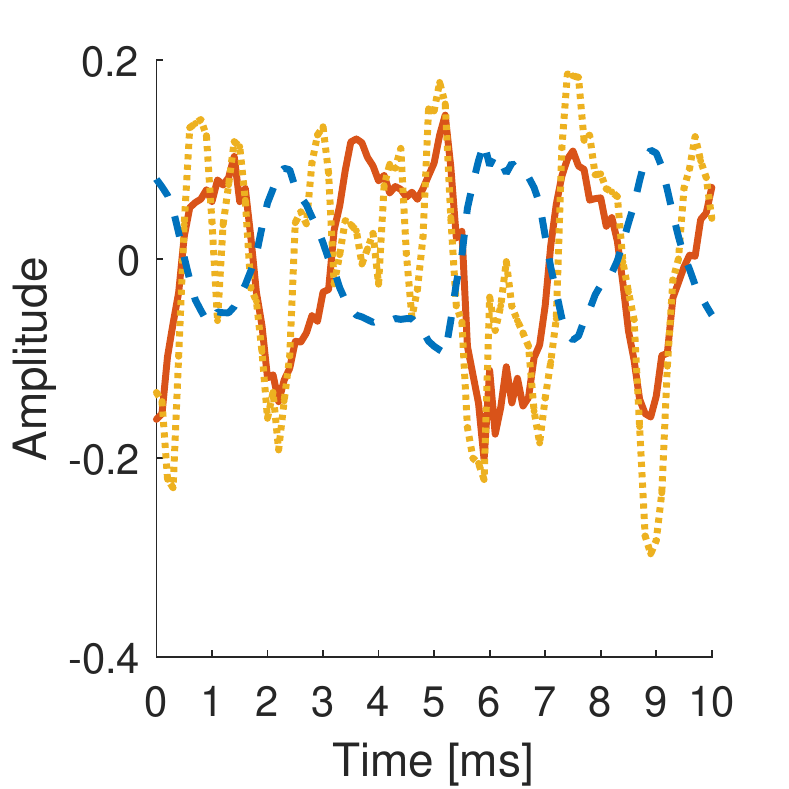}}
		\subcaption{$\mathcal{L}_{\text{SI-SDR}}$}
		\label{fig:fig4104}
	\end{minipage}
	\begin{minipage}{.49\columnwidth}
		\centering
		\centerline{\includegraphics[trim={0mm 0mm 0mm 0mm},clip,width=0.95\linewidth]{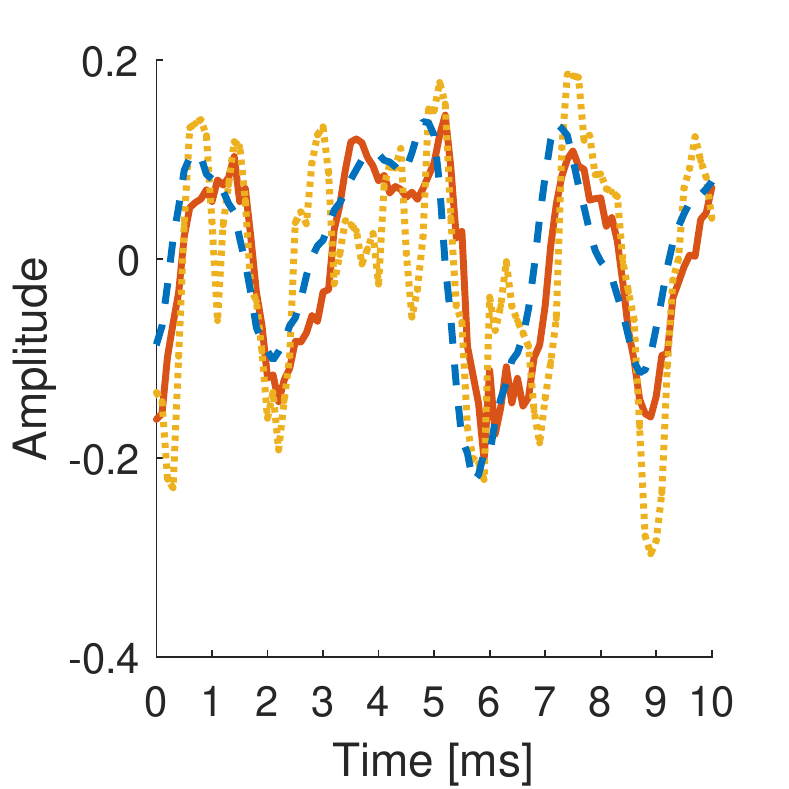}}
		\subcaption{$\mathcal{L}_{\text{STSA-MSE}}$}
		\label{fig:fig5104}
	\end{minipage}%
	\begin{minipage}{.49\columnwidth}
		\centering
		\centerline{\includegraphics[trim={0mm 0mm 0mm 0mm},clip,width=0.95\linewidth]{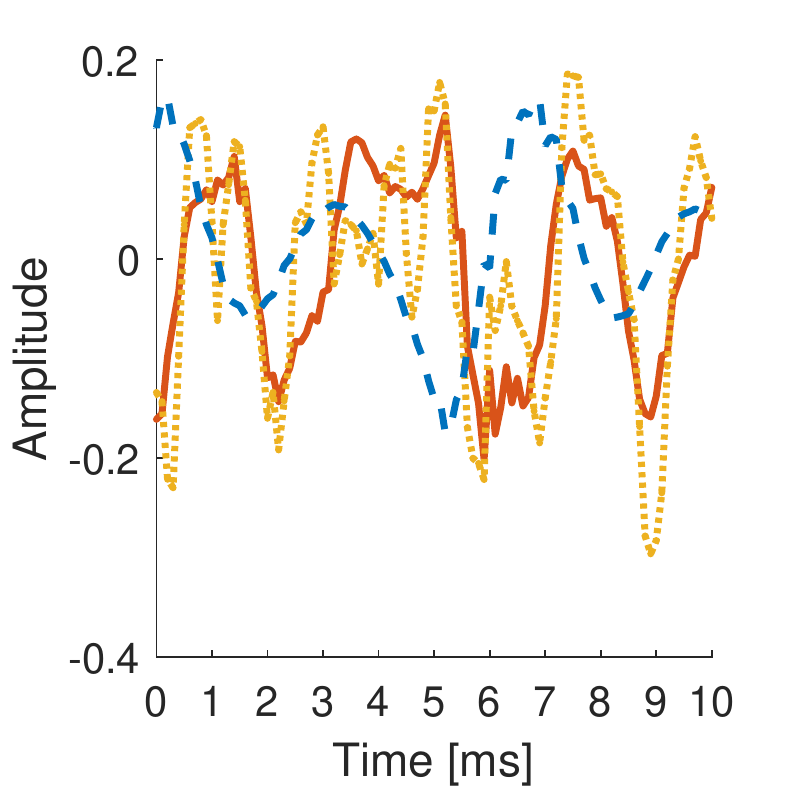}}
		\subcaption{$\mathcal{L}_{\text{PMSQE}}$}
		\label{fig:fig7402}
	\end{minipage}
	\caption{Time-domain waveform of a clean speech signal (solid red), noisy speech signal (dotted yellow) and processed speech signals (dashed blue) processed by systems trained using different loss functions.  }
	\label{fig:lr}
\end{figure}
\begin{figure}
	\centering
	\begin{minipage}{.49\columnwidth}
		\centering
		\centerline{\includegraphics[trim={0mm 0mm 0mm 0mm},clip,width=0.95\linewidth]{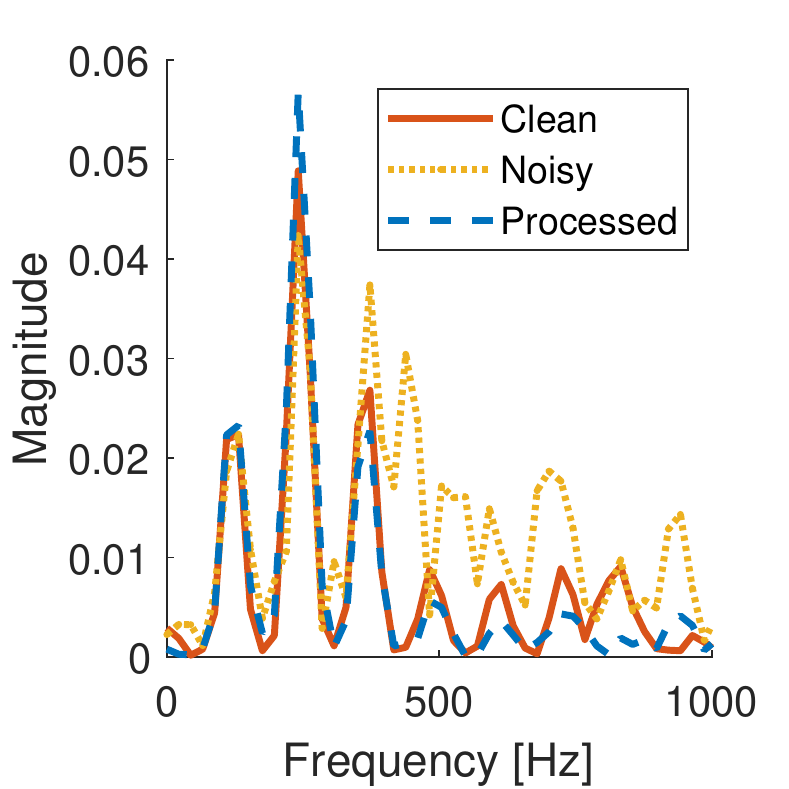}}
		\subcaption{$\mathcal{L}_{\text{TIME-MSE}}$}
		\label{fig:fig1104_freq}
	\end{minipage}%
	\begin{minipage}{.49\columnwidth}
		\centering
		\centerline{\includegraphics[trim={0mm 0mm 0mm 0mm},clip,width=0.95\linewidth]{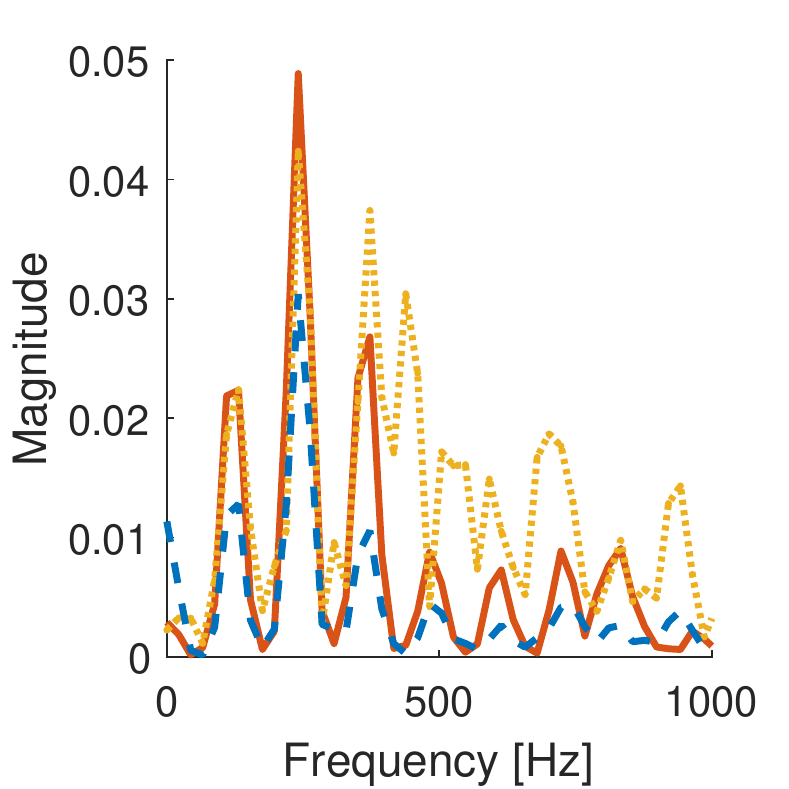}}
		\label{fig:fig2104_freq}
		\subcaption{$\mathcal{L}_{\text{STOI}}$}
	\end{minipage}
	\begin{minipage}{.49\columnwidth}
		\centering
		\centerline{\includegraphics[trim={0mm 0mm 0mm 0mm},clip,width=0.95\linewidth]{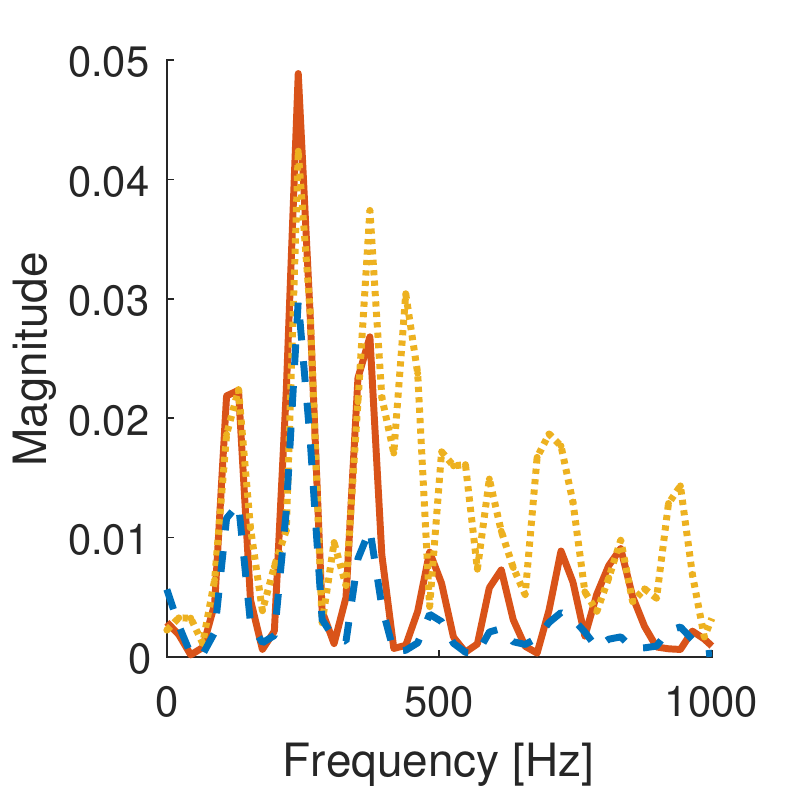}}
		\label{fig:fig3104_freq}
		\subcaption{$\mathcal{L}_{\text{ESTOI}}$}
	\end{minipage}%
	\begin{minipage}{.49\columnwidth}
		\centering
		\centerline{\includegraphics[trim={0mm 0mm 0mm 0mm},clip,width=0.95\linewidth]{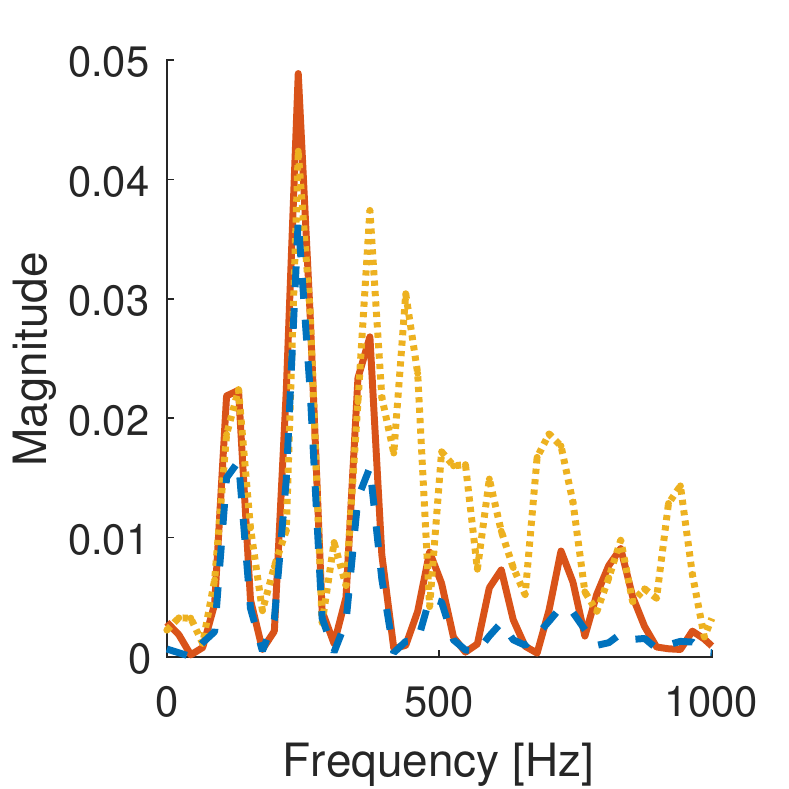}}
		\label{fig:fig4104_freq}
		\subcaption{$\mathcal{L}_{\text{SI-SDR}}$}
	\end{minipage}
	\begin{minipage}{.49\columnwidth}
		\centering
		\centerline{\includegraphics[trim={0mm 0mm 0mm 0mm},clip,width=0.95\linewidth]{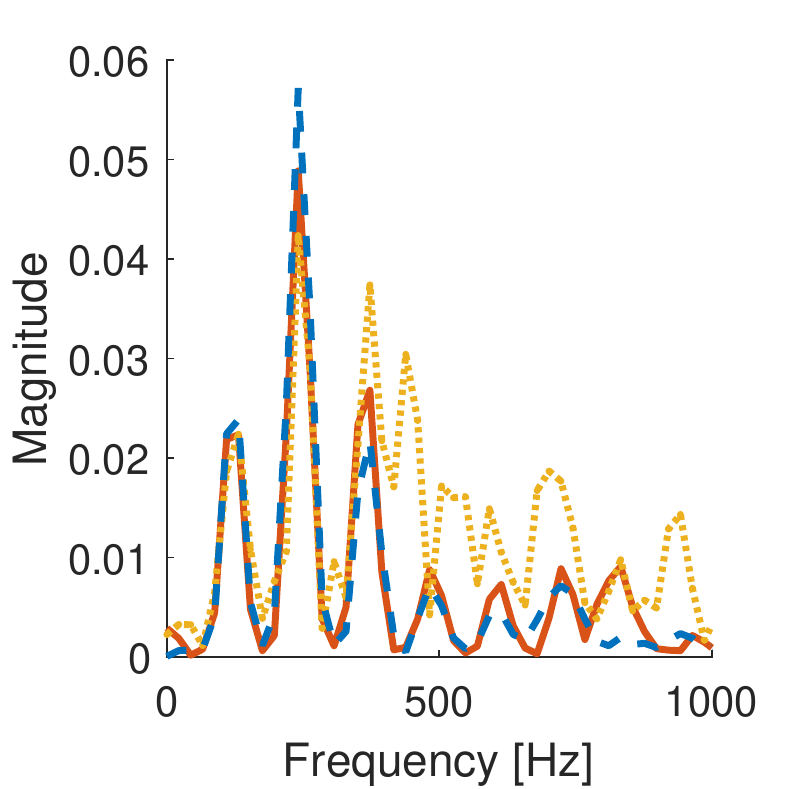}}
		\subcaption{$\mathcal{L}_{\text{STSA-MSE}}$}
		\label{fig:fig5104_freq}
	\end{minipage}%
	\begin{minipage}{.49\columnwidth}
		\centering
		\centerline{\includegraphics[trim={0mm 0mm 0mm 0mm},clip,width=0.95\linewidth]{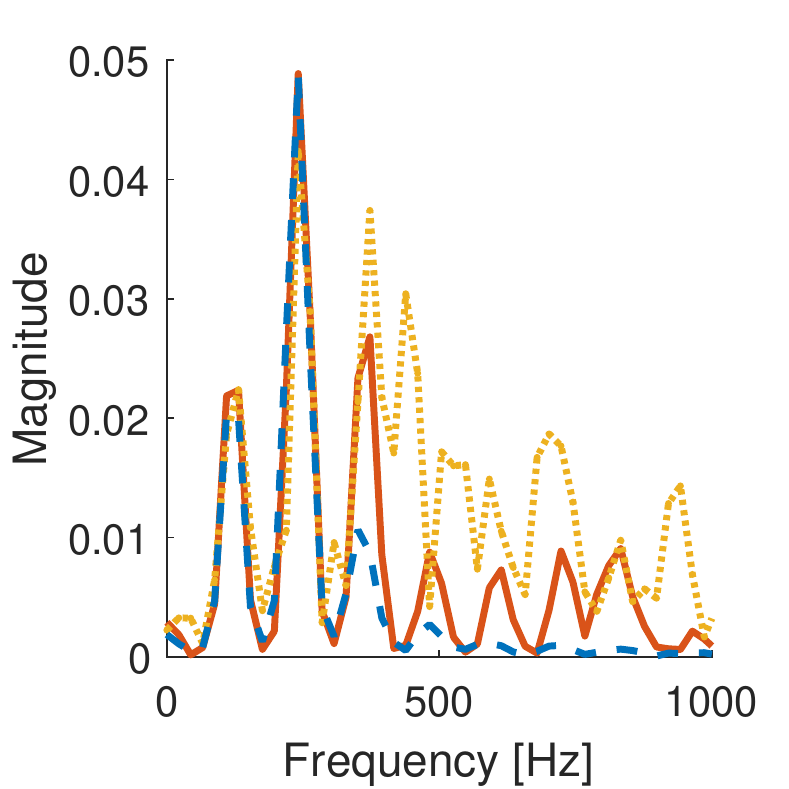}}
		\label{fig:fig7402_freq}
		\subcaption{$\mathcal{L}_{\text{PMSQE}}$}
	\end{minipage}
	\caption{Magnitude spectra of the signals presented in Fig.\,\ref{fig:lr}.}
	\label{fig:lr_freq}
\end{figure}
\subsection{Signal Integrity vs. Performance Metric}\label{sec:exresSI}
We now study the signal integrity achieved by the systems trained to minimize the different loss functions. Specifically, we compare the waveforms (Fig.\;\ref{fig:lr}) and amplitude spectra (Fig.\;\ref{fig:lr_freq}) of representative clean, noisy, and enhanced speech signal segments processed by speech enhancement systems trained using the loss functions presented in Sec.\,\ref{sec:sesys} and the learning rates given in Table\,\ref{tab:models}.  

Figure\;\ref{fig:lr} presents the waveforms of a specific 10 ms realization of clean, noisy, and enhanced speech signals from the experiments in Table\;\ref{tab:lr_res}.  
At first, if polarity is ignored, it is seen from Fig.\;\ref{fig:lr} that systems trained with the six loss functions manage to enhance the noisy speech signal, as we see a somewhat good correspondence between the clean speech signal (red-solid) and the enhanced speech signal (blue-dashed) with the enhanced signal having considerably less noise compared to the noisy speech signal (yellow-dotted). 
From Fig.\,\ref{fig:fig1104} it appears that the $\mathcal{L}_{\text{TIME-MSE}}$ loss function achieves the most per-sample-accurate estimate of the clean signal and $\mathcal{L}_{\text{STOI}}$, $\mathcal{L}_{\text{ESTOI}}$, and $\mathcal{L}_{\text{PMSQE}}$ (Figs.\,\ref{fig:fig2104},\,\ref{fig:fig3104}, and\,\ref{fig:fig7402}) appear to achieve the least per-sample-accurate estimate. 
It is not surprising that $\mathcal{L}_{\text{TIME-MSE}}$ achieves the best estimate, as $\mathcal{L}_{\text{TIME-MSE}}$ is a waveform matching loss function and consequently penalizes time-domain samples that deviate from the samples of the clean signal. 
However, a perfect sample-wise waveform reconstruction is not necessarily the only optimum, if the receiver is the human auditory system and the goal is to achieve high speech intelligibility or quality as perceived by humans. For example, in Figs.\,\ref{fig:fig2104} and \ref{fig:fig3104} it is seen that the waveforms of the enhanced signals are inverted, and somewhat different from Fig.\,\ref{fig:fig1104} although the processed signals achieve similar or higher STOI and ESTOI scores (Table\;\ref{tab:lr_res}), i.e. the signals should ideally represent similar or higher levels of intelligibility. 
This is because $\mathcal{L}_{\text{STOI}}$ and $\mathcal{L}_{\text{ESTOI}}$ are loss functions based on matching of short-time energy in one-third octave bands. As a consequence, these loss functions are e.g. invariant to the signal polarity.

Furthermore, by careful inspecting e.g. Fig.\,\ref{fig:fig3104} it can be observed that the enhanced signal is slightly time-shifted compared to the clean signal. This phenomenon is more evident in Fig.\,\ref{fig:fig5104}, where it is easy to see that the enhanced signal is time-shifted a few samples with respect to the clean signal. 
Loss functions such as $\mathcal{L}_{\text{STOI}}$, $\mathcal{L}_{\text{ESTOI}}$, $\mathcal{L}_{\text{PMSQE}}$ and $\mathcal{L}_{\text{STSA-MSE}}$ are primarily based on short-time magnitude spectra, and do not, penalize waveform deviations. Hence, they may allow for the enhanced speech signal to be time-shifted with respect to the clean signal. 
That being said, the amount of time-shift that we have observed is less than 1 ms, which is considerably smaller than the 10-30 ms usually required before the time-shift may be perceivable in real-life low-latency speech processing applications such as hearing aids and mobile communications devices \cite{stone_tolerable_1999,bramslow_preferred_2010}.

Furthermore, when evaluating speech enhancement performance with waveform-matching metrics such as SNR, SDR, or SI-SDR, exact time-matching is critical and a few samples delay can cause a complete failure of such performance metrics.   
This is exactly what we observe in Table\;\ref{tab:lr_res}, where the SI-SDR scores, and to a smaller extent the SDR scores, have a high variance with no obvious correspondence with the remaining metrics such as STOI and ESTOI, which are stable and show a more consistent behavior. Since SI-SDR and SDR are scale-invariant waveform matching functions, they fail if the processed signal and the reference signal are not perfectly aligned. 
Consequently, SI-SDR and SDR should be used with caution, when they are used to evaluate time-domain speech enhancement or separation systems with the capability to modify the phase, such as time-domain FCNNs.
Furthermore, they should generally be avoided, when loss functions like $\mathcal{L}_{\text{STOI}}$, $\mathcal{L}_{\text{ESTOI}}$, $\mathcal{L}_{\text{PMSQE}}$ and $\mathcal{L}_{\text{STSA-MSE}}$ are utilized. 

Finally, in Fig.\;\ref{fig:lr_freq} we show the corresponding amplitude spectra of the signals from Fig.\;\ref{fig:lr} using a 40 ms window, centered around the 10 ms time-domain segment from Fig.\;\ref{fig:lr}, to ensure a sufficient frequency resolution. 
It is seen from Fig.\;\ref{fig:lr_freq} that all six loss functions lead to enhanced signals whose magnitude spectrum resemble that of the magnitude spectrum of the clean speech signal. Furthermore, it is seen that the enhanced signals capture the dominating harmonics of the clean speech signal, while attenuating the major frequency components of the noise.
Similarly to Fig.\,\ref{fig:fig1104}, it is seen that $\mathcal{L}_{\text{TIME-MSE}}$ achieves an accurate estimate of the amplitude spectrum but also $\mathcal{L}_{\text{STSA-MSE}}$ (Fig.\,\ref{fig:fig1104_freq}) achieves an accurate estimate, which is expected as $\mathcal{L}_{\text{STSA-MSE}}$ is a frequency-domain energy-matching loss function. 
In fact, by careful inspection of Fig.\,\ref{fig:fig5104_freq} it can be observed that $\mathcal{L}_{\text{STSA-MSE}}$ manages to preserve the higher order harmonics to a larger extent than $\mathcal{L}_{\text{TIME-MSE}}$ (Fig.\,\ref{fig:fig1104_freq}).  
Also, as expected, we can conclude that the small time-shift induced by $\mathcal{L}_{\text{STSA-MSE}}$ in Fig.\,\ref{fig:fig5104} has no apparent effect on the accuracy of the amplitude spectrum estimate, which indicates that the time-shift is approximately constant over the window length of $40$ ms.

\begin{table*}
	\caption{Performance of different speech enhancement systems measured using STOI, ESTOI, SI-SDR, and SDR. The systems have been trained using different loss functions ($\mathcal{L}_{\text{TIME-MSE}}$, $\mathcal{L}_{\text{STOI}}$, $\mathcal{L}_{\text{ESTOI}}$, $\mathcal{L}_{\text{SI-SDR}}$, $\mathcal{L}_{\text{STSA-MSE}}$, $\mathcal{L}_{\text{TIME-MSE}}$) and tested using four matched (SSN, BBL, CAF, STR) noise types at seven different SNRs (-10 dB, -5 dB, 0 dB, 5 dB, 10 dB, 15 dB, and 20dB). The maximum score is highlighted in boldface for each SNR and performance measure. See text for details.}\label{tab:nt_res}
	\begin{subtable}{.5\textwidth}
		\caption{Speech Shaped Noise (matched)}\label{tab:nt_res_ssn}
		\centering
		\setlength\tabcolsep{5pt} 
		\resizebox{0.99\columnwidth}{!}{%
			\begin{tabular}{cllcccccc}
				\midrule
			    \multirow{2}{*}{SNR} & \multirow{2}{*}{Metric} & \multirow{2}{*}{Noisy} &\multicolumn{6}{c}{Processed} \\ \cmidrule(lr){4-9}
				&  &  &$\mathcal{L}_{\text{TIME-MSE}}$ & $\mathcal{L}_{\text{STOI}}$ & $\mathcal{L}_{\text{ESTOI}}$ & $\mathcal{L}_{\text{SI-SDR}}$ & $\mathcal{L}_{\text{STSA-MSE}}$  & $\mathcal{L}_{\text{PMSQE}}$  \\	
				\midrule 
				\multirow{ 5}{*}{-10 dB:} 
& STOI:  & 0.50 & 0.63 & \textbf{0.68} & 0.64 & 0.64 & 0.63 & 0.56\\ 
& ESTOI: & 0.16 & 0.34 & 0.40 & \textbf{0.43} & 0.36 & 0.35 & 0.27\\ 
& SI-SDR:& -11.07 & 0.41 & -7.34 & -16.18 & \textbf{0.67} & -6.93 & -24.50\\ 
& SDR:   & -10.32 & 1.84 & -0.76 & -1.63 & \textbf{2.05} & 0.66 & -1.52\\ 
& PESQ:  & 1.48 & 1.59 & 1.60 & 1.31 & 1.62 & \textbf{1.68} & 1.47\\ 
\midrule 
\multirow{ 5}{*}{-5 dB:} 
& STOI:  & 0.62 & 0.82 & \textbf{0.85} & 0.83 & 0.83 & 0.82 & 0.77\\ 
& ESTOI: & 0.30 & 0.60 & 0.65 & \textbf{0.67} & 0.62 & 0.60 & 0.54\\ 
& SI-SDR:& -6.05 & 5.74 & -3.53 & -13.41 & \textbf{6.08} & -3.74 & -20.08\\ 
& SDR:   & -5.77 & 6.60 & 4.21 & 3.63 & \textbf{6.87} & 5.48 & 2.37\\ 
& PESQ:  & 1.59 & 2.19 & 2.18 & 2.03 & 2.22 & \textbf{2.24} & 2.11\\ 
\midrule 
\multirow{ 5}{*}{0 dB:} 
& STOI:  & 0.75 & 0.90 & \textbf{0.92} & \textbf{0.92} & 0.91 & 0.90 & 0.88\\ 
& ESTOI: & 0.46 & 0.75 & 0.80 & \textbf{0.81} & 0.78 & 0.75 & 0.71\\ 
& SI-SDR:& -1.05 & 9.46 & -2.16 & -12.73 & \textbf{9.96} & -2.66 & -17.95\\ 
& SDR:   & -0.92 & 10.09 & 7.54 & 6.73 & \textbf{10.55} & 8.75 & 4.69\\ 
& PESQ:  & 1.79 & 2.59 & 2.58 & 2.56 & \textbf{2.65} & 2.62 & 2.57\\ 
\midrule 
\multirow{ 5}{*}{5 dB:} 
& STOI:  & 0.85 & 0.94 & \textbf{0.95} & \textbf{0.95} & \textbf{0.95} & 0.94 & 0.92\\ 
& ESTOI: & 0.63 & 0.84 & 0.87 & \textbf{0.88} & 0.86 & 0.84 & 0.80\\ 
& SI-SDR:& 3.96 & 12.56 & -1.51 & -12.46 & \textbf{13.20} & -2.12 & -17.23\\ 
& SDR:   & 4.04 & 13.07 & 10.00 & 8.68 & \textbf{13.71} & 11.30 & 6.07\\ 
& PESQ:  & 2.03 & 2.89 & 2.88 & 2.91 & \textbf{3.00} & 2.89 & 2.89\\ 
\midrule 
\multirow{ 5}{*}{10 dB:} 
& STOI:  & 0.92 & 0.96 & \textbf{0.97} & \textbf{0.97} & \textbf{0.97} & 0.96 & 0.94\\ 
& ESTOI: & 0.78 & 0.90 & 0.91 & \textbf{0.92} & 0.91 & 0.90 & 0.85\\ 
& SI-SDR:& 8.96 & 15.39 & -1.18 & -12.25 & \textbf{16.10} & -1.86 & -16.98\\ 
& SDR:   & 9.02 & 15.83 & 11.84 & 9.98 & \textbf{16.60} & 13.39 & 6.85\\ 
& PESQ:  & 2.32 & 3.13 & 3.12 & 3.16 & \textbf{3.27} & 3.10 & 3.13\\ 
\midrule 
\multirow{ 5}{*}{15 dB:} 
& STOI:  & 0.96 & 0.97 & \textbf{0.98} & \textbf{0.98} & \textbf{0.98} & 0.97 & 0.95\\ 
& ESTOI: & 0.88 & 0.93 & \textbf{0.94} & \textbf{0.94} & \textbf{0.94} & 0.93 & 0.88\\ 
& SI-SDR:& 13.96 & 17.89 & -1.02 & -12.11 & \textbf{18.71} & -1.74 & -16.95\\ 
& SDR:   & 14.02 & 18.36 & 13.06 & 10.76 & \textbf{19.31} & 14.94 & 7.26\\ 
& PESQ:  & 2.63 & 3.31 & 3.33 & 3.36 & \textbf{3.47} & 3.29 & 3.32\\ 
\midrule 
\multirow{ 5}{*}{20 dB:} 
& STOI:  & 0.98 & 0.98 & 0.98 & 0.98 & \textbf{0.99} & 0.98 & 0.96\\ 
& ESTOI: & 0.94 & 0.94 & 0.95 & 0.95 & \textbf{0.96} & 0.94 & 0.90\\ 
& SI-SDR:& 18.96 & 19.73 & -0.95 & -12.05 & \textbf{20.89} & -1.68 & -16.97\\ 
& SDR:   & 19.02 & 20.29 & 13.75 & 11.17 & \textbf{21.71} & 15.80 & 7.45\\ 
& PESQ:  & 2.95 & 3.48 & 3.55 & 3.54 & 3.63 & 3.50 & 3.46\\ 
\midrule 
 
		\end{tabular}}
	\end{subtable}
	\begin{subtable}{.5\textwidth}
		\caption{6-Speaker Babble Noise (matched)}\label{tab:nt_res_bbl}
		\centering
		\setlength\tabcolsep{5pt} 
		\resizebox{0.99\columnwidth}{!}{%
			\begin{tabular}{cllcccccc}
				\midrule
				\multirow{2}{*}{SNR} & \multirow{2}{*}{Metric} & \multirow{2}{*}{Noisy} &\multicolumn{6}{c}{Processed} \\ \cmidrule(lr){4-9}
				&  &  &$\mathcal{L}_{\text{TIME-MSE}}$ & $\mathcal{L}_{\text{STOI}}$ & $\mathcal{L}_{\text{ESTOI}}$ & $\mathcal{L}_{\text{SI-SDR}}$ & $\mathcal{L}_{\text{STSA-MSE}}$  & $\mathcal{L}_{\text{PMSQE}}$  \\	
				\midrule 
				\multirow{ 5}{*}{-10 dB:} 
& STOI:  & 0.46 & 0.67 & \textbf{0.69} & 0.66 & 0.68 & 0.67 & 0.59\\ 
& ESTOI: & 0.18 & 0.41 & 0.43 & \textbf{0.44} & 0.42 & 0.41 & 0.33\\ 
& SI-SDR:& -11.04 & 0.39 & -6.70 & -15.84 & \textbf{0.59} & -6.49 & -25.93\\ 
& SDR:   & -10.29 & 1.69 & -0.51 & -1.38 & \textbf{1.86} & 0.57 & -1.67\\ 
& PESQ:  & 1.73 & 1.70 & 1.62 & 1.41 & 1.70 & \textbf{1.72} & 1.51\\ 
\midrule 
\multirow{ 5}{*}{-5 dB:} 
& STOI:  & 0.59 & 0.82 & \textbf{0.84} & 0.83 & 0.83 & 0.82 & 0.76\\ 
& ESTOI: & 0.31 & 0.62 & 0.65 & \textbf{0.66} & 0.64 & 0.61 & 0.55\\ 
& SI-SDR:& -6.04 & 5.59 & -3.29 & -13.54 & \textbf{5.96} & -3.69 & -20.38\\ 
& SDR:   & -5.75 & 6.45 & 4.36 & 3.66 & \textbf{6.75} & 5.32 & 2.24\\ 
& PESQ:  & 1.69 & 2.18 & 2.16 & 2.02 & \textbf{2.21} & 2.18 & 2.04\\ 
\midrule 
\multirow{ 5}{*}{0 dB:} 
& STOI:  & 0.73 & 0.90 & \textbf{0.92} & 0.91 & 0.91 & 0.90 & 0.87\\ 
& ESTOI: & 0.47 & 0.76 & 0.80 & \textbf{0.81} & 0.78 & 0.76 & 0.71\\ 
& SI-SDR:& -1.04 & 9.54 & -1.94 & -12.81 & \textbf{10.12} & -2.48 & -18.13\\ 
& SDR:   & -0.91 & 10.17 & 7.78 & 6.82 & \textbf{10.69} & 8.83 & 4.58\\ 
& PESQ:  & 1.84 & 2.57 & 2.60 & 2.54 & \textbf{2.64} & 2.57 & 2.49\\ 
\midrule 
\multirow{ 5}{*}{5 dB:} 
& STOI:  & 0.84 & 0.94 & \textbf{0.95} & \textbf{0.95} & \textbf{0.95} & 0.94 & 0.92\\ 
& ESTOI: & 0.63 & 0.85 & 0.87 & \textbf{0.88} & 0.87 & 0.84 & 0.80\\ 
& SI-SDR:& 3.96 & 12.75 & -1.40 & -12.55 & \textbf{13.49} & -2.03 & -17.25\\ 
& SDR:   & 4.04 & 13.26 & 10.20 & 8.76 & \textbf{13.99} & 11.46 & 5.99\\ 
& PESQ:  & 2.10 & 2.88 & 2.92 & 2.90 & \textbf{2.98} & 2.86 & 2.84\\ 
\midrule 
\multirow{ 5}{*}{10 dB:} 
& STOI:  & 0.91 & 0.96 & \textbf{0.97} & \textbf{0.97} & \textbf{0.97} & 0.96 & 0.94\\ 
& ESTOI: & 0.77 & 0.90 & \textbf{0.92} & \textbf{0.92} & \textbf{0.92} & 0.89 & 0.85\\ 
& SI-SDR:& 8.96 & 15.55 & -1.15 & -12.31 & \textbf{16.41} & -1.82 & -16.97\\ 
& SDR:   & 9.03 & 16.02 & 11.94 & 10.00 & \textbf{16.91} & 13.49 & 6.79\\ 
& PESQ:  & 2.39 & 3.12 & 3.17 & 3.17 & \textbf{3.25} & 3.10 & 3.11\\ 
\midrule 
\multirow{ 5}{*}{15 dB:} 
& STOI:  & 0.96 & 0.97 & \textbf{0.98} & \textbf{0.98} & \textbf{0.98} & 0.97 & 0.95\\ 
& ESTOI: & 0.86 & 0.93 & \textbf{0.94} & \textbf{0.94} & \textbf{0.94} & 0.92 & 0.88\\ 
& SI-SDR:& 13.96 & 17.96 & -0.99 & -12.11 & \textbf{18.96} & -1.72 & -16.94\\ 
& SDR:   & 14.02 & 18.46 & 13.09 & 10.74 & \textbf{19.56} & 14.95 & 7.20\\ 
& PESQ:  & 2.70 & 3.33 & 3.38 & 3.38 & \textbf{3.46} & 3.32 & 3.30\\ 
\midrule 
\multirow{ 5}{*}{20 dB:} 
& STOI:  & 0.98 & 0.98 & 0.98 & 0.98 & \textbf{0.99} & 0.98 & 0.96\\ 
& ESTOI: & 0.93 & 0.94 & 0.95 & 0.95 & \textbf{0.96} & 0.94 & 0.89\\ 
& SI-SDR:& 18.96 & 19.70 & -0.94 & -12.06 & \textbf{21.03} & -1.68 & -16.94\\ 
& SDR:   & 19.02 & 20.28 & 13.75 & 11.15 & \textbf{21.86} & 15.80 & 7.40\\ 
& PESQ:  & 3.02 & 3.50 & 3.57 & 3.54 & \textbf{3.62} & 3.52 & 3.44\\ 
\midrule 
 
		\end{tabular}}
	\end{subtable}
\newline
\vspace*{5 mm}
\newline
\begin{subtable}{.5\textwidth}
		\caption{Cafeteria Noise (matched)}\label{tab:nt_res_caf}
		\centering
		\setlength\tabcolsep{5pt} 
		\resizebox{0.99\columnwidth}{!}{%
			\begin{tabular}{cllcccccc}
				\midrule
				\multirow{2}{*}{SNR} & \multirow{2}{*}{Metric} & \multirow{2}{*}{Noisy} &\multicolumn{6}{c}{Processed} \\ \cmidrule(lr){4-9}
				&  &  &$\mathcal{L}_{\text{TIME-MSE}}$ & $\mathcal{L}_{\text{STOI}}$ & $\mathcal{L}_{\text{ESTOI}}$ & $\mathcal{L}_{\text{SI-SDR}}$ & $\mathcal{L}_{\text{STSA-MSE}}$  & $\mathcal{L}_{\text{PMSQE}}$  \\	
				\multirow{ 5}{*}{-10 dB:} 
& STOI:  & 0.56 & 0.74 & \textbf{0.77} & 0.74 & 0.74 & 0.74 & 0.69\\ 
& ESTOI: & 0.25 & 0.48 & 0.52 & \textbf{0.53} & 0.49 & 0.48 & 0.42\\ 
& SI-SDR:& -11.04 & 3.26 & -5.16 & -14.90 & \textbf{3.78} & -5.37 & -19.30\\ 
& SDR:   & -10.32 & 4.21 & 1.95 & 1.23 & \textbf{4.74} & 2.90 & 0.41\\ 
& PESQ:  & 1.60 & 1.99 & 1.97 & 1.80 & \textbf{2.04} & 2.03 & 1.87\\ 
\midrule 
\multirow{ 5}{*}{-5 dB:} 
& STOI:  & 0.68 & 0.85 & \textbf{0.87} & 0.86 & 0.86 & 0.86 & 0.82\\ 
& ESTOI: & 0.38 & 0.65 & 0.70 & \textbf{0.71} & 0.67 & 0.66 & 0.60\\ 
& SI-SDR:& -6.03 & 7.77 & -2.69 & -13.38 & \textbf{8.33} & -3.20 & -17.83\\ 
& SDR:   & -5.75 & 8.49 & 6.22 & 5.34 & \textbf{9.06} & 7.14 & 3.57\\ 
& PESQ:  & 1.69 & 2.39 & 2.39 & 2.29 & \textbf{2.45} & 2.41 & 2.29\\ 
\midrule 
\multirow{ 5}{*}{0 dB:} 
& STOI:  & 0.78 & 0.92 & \textbf{0.93} & \textbf{0.93} & 0.92 & 0.92 & 0.89\\ 
& ESTOI: & 0.52 & 0.77 & 0.81 & \textbf{0.82} & 0.80 & 0.78 & 0.73\\ 
& SI-SDR:& -1.03 & 11.17 & -1.69 & -12.83 & \textbf{11.88} & -2.35 & -17.30\\ 
& SDR:   & -0.90 & 11.75 & 9.22 & 8.01 & \textbf{12.50} & 10.18 & 5.54\\ 
& PESQ:  & 1.99 & 2.70 & 2.73 & 2.69 & \textbf{2.79} & 2.72 & 2.65\\ 
\midrule 
\multirow{ 5}{*}{5 dB:} 
& STOI:  & 0.87 & 0.95 & \textbf{0.96} & \textbf{0.96} & 0.95 & 0.95 & 0.93\\ 
& ESTOI: & 0.66 & 0.85 & \textbf{0.88} & \textbf{0.88} & 0.87 & 0.85 & 0.81\\ 
& SI-SDR:& 3.97 & 14.11 & -1.28 & -12.42 & \textbf{14.90} & -1.98 & -17.10\\ 
& SDR:   & 4.05 & 14.61 & 11.32 & 9.67 & \textbf{15.50} & 12.55 & 6.66\\ 
& PESQ:  & 2.33 & 2.96 & 3.01 & 3.01 & \textbf{3.08} & 2.97 & 2.96\\ 
\midrule 
\multirow{ 5}{*}{10 dB:} 
& STOI:  & 0.93 & \textbf{0.97} & \textbf{0.97} & \textbf{0.97} & \textbf{0.97} & \textbf{0.97} & 0.95\\ 
& ESTOI: & 0.79 & 0.90 & \textbf{0.92} & \textbf{0.92} & \textbf{0.92} & 0.90 & 0.86\\ 
& SI-SDR:& 8.97 & 16.78 & -1.09 & -12.18 & \textbf{17.58} & -1.81 & -17.03\\ 
& SDR:   & 9.03 & 17.28 & 12.76 & 10.66 & \textbf{18.27} & 14.38 & 7.22\\ 
& PESQ:  & 2.66 & 3.20 & 3.27 & 3.27 & 3.32 & 3.21 & 3.21\\ 
\midrule 
\multirow{ 5}{*}{15 dB:} 
& STOI:  & 0.96 & \textbf{0.98} & \textbf{0.98} & \textbf{0.98} & \textbf{0.98} & \textbf{0.98} & 0.95\\ 
& ESTOI: & 0.88 & 0.93 & \textbf{0.94} & \textbf{0.94} & \textbf{0.94} & 0.93 & 0.89\\ 
& SI-SDR:& 13.97 & 18.91 & -1.00 & -12.06 & \textbf{19.94} & -1.72 & -17.02\\ 
& SDR:   & 14.03 & 19.48 & 13.61 & 11.18 & \textbf{20.84} & 15.55 & 7.47\\ 
& PESQ:  & 2.99 & 3.42 & 3.50 & 3.49 & 3.53 & 3.44 & 3.40\\ 
\midrule 
\multirow{ 5}{*}{20 dB:} 
& STOI:  & 0.98 & 0.98 & 0.98 & 0.98 & \textbf{0.99} & 0.98 & 0.96\\ 
& ESTOI: & 0.94 & 0.94 & 0.95 & 0.95 & \textbf{0.96} & 0.94 & 0.90\\ 
& SI-SDR:& 18.97 & 20.22 & -0.96 & -12.02 & \textbf{21.75} & -1.68 & -17.01\\ 
& SDR:   & 19.02 & 20.85 & 14.01 & 11.40 & \textbf{22.91} & 16.10 & 7.55\\ 
& PESQ:  & 3.33 & 3.60 & 3.70 & 3.67 & \textbf{3.72} & 3.63 & 3.52\\ 
\midrule 
 
		\end{tabular}}
	\end{subtable}
	\begin{subtable}{.5\textwidth}
		\caption{Street Noise (matched)}\label{tab:nt_res_str}
		\centering
		\setlength\tabcolsep{5pt} 
		\resizebox{0.99\columnwidth}{!}{%
			\begin{tabular}{cllcccccc}
				\midrule
				\multirow{2}{*}{SNR} & \multirow{2}{*}{Metric} & \multirow{2}{*}{Noisy} &\multicolumn{6}{c}{Processed} \\ \cmidrule(lr){4-9}
				&  &  &$\mathcal{L}_{\text{TIME-MSE}}$ & $\mathcal{L}_{\text{STOI}}$ & $\mathcal{L}_{\text{ESTOI}}$ & $\mathcal{L}_{\text{SI-SDR}}$ & $\mathcal{L}_{\text{STSA-MSE}}$  & $\mathcal{L}_{\text{PMSQE}}$  \\	
				\multirow{ 5}{*}{-10 dB:} 
& STOI:  & 0.58 & 0.76 & \textbf{0.80} & 0.77 & 0.77 & 0.76 & 0.70\\ 
& ESTOI: & 0.24 & 0.50 & 0.56 & \textbf{0.57} & 0.52 & 0.50 & 0.43\\ 
& SI-SDR:& -11.05 & 3.92 & -4.74 & -15.07 & \textbf{4.34} & -5.07 & -18.52\\ 
& SDR:   & -10.35 & 4.92 & 2.64 & 1.79 & \textbf{5.32} & 3.53 & 0.46\\ 
& PESQ:  & 1.42 & 2.05 & 2.02 & 1.84 & \textbf{2.07} & \textbf{2.07} & 1.90\\ 
\midrule 
\multirow{ 5}{*}{-5 dB:} 
& STOI:  & 0.68 & 0.87 & \textbf{0.89} & 0.88 & 0.88 & 0.87 & 0.83\\ 
& ESTOI: & 0.36 & 0.68 & 0.72 & \textbf{0.73} & 0.69 & 0.68 & 0.61\\ 
& SI-SDR:& -6.04 & 8.12 & -2.58 & -13.50 & \textbf{8.58} & -3.24 & -17.63\\ 
& SDR:   & -5.77 & 8.88 & 6.60 & 5.67 & \textbf{9.32} & 7.46 & 3.46\\ 
& PESQ:  & 1.63 & 2.45 & 2.44 & 2.36 & \textbf{2.50} & 2.46 & 2.35\\ 
\midrule 
\multirow{ 5}{*}{0 dB:} 
& STOI:  & 0.78 & 0.92 & \textbf{0.94} & 0.93 & 0.93 & 0.92 & 0.90\\ 
& ESTOI: & 0.49 & 0.79 & 0.82 & \textbf{0.83} & 0.81 & 0.79 & 0.74\\ 
& SI-SDR:& -1.04 & 11.39 & -1.66 & -12.81 & \textbf{11.97} & -2.38 & -17.32\\ 
& SDR:   & -0.92 & 12.03 & 9.46 & 8.20 & \textbf{12.62} & 10.36 & 5.46\\ 
& PESQ:  & 1.94 & 2.76 & 2.78 & 2.75 & \textbf{2.84} & 2.76 & 2.71\\ 
\midrule 
\multirow{ 5}{*}{5 dB:} 
& STOI:  & 0.86 & 0.95 & \textbf{0.96} & \textbf{0.96} & \textbf{0.96} & 0.95 & 0.93\\ 
& ESTOI: & 0.63 & 0.86 & 0.88 & \textbf{0.89} & 0.87 & 0.86 & 0.82\\ 
& SI-SDR:& 3.96 & 14.28 & -1.26 & -12.45 & \textbf{14.90} & -1.99 & -17.16\\ 
& SDR:   & 4.04 & 14.83 & 11.53 & 9.79 & \textbf{15.54} & 12.69 & 6.66\\ 
& PESQ:  & 2.28 & 3.01 & 3.05 & 3.04 & \textbf{3.12} & 2.99 & 3.00\\ 
\midrule 
\multirow{ 5}{*}{10 dB:} 
& STOI:  & 0.92 & \textbf{0.97} & \textbf{0.97} & \textbf{0.97} & \textbf{0.97} & \textbf{0.97} & 0.95\\ 
& ESTOI: & 0.76 & 0.90 & \textbf{0.92} & \textbf{0.92} & \textbf{0.92} & 0.90 & 0.87\\ 
& SI-SDR:& 8.96 & 16.96 & -1.09 & -12.17 & \textbf{17.60} & -1.80 & -17.06\\ 
& SDR:   & 9.02 & 17.51 & 12.92 & 10.75 & \textbf{18.34} & 14.53 & 7.27\\ 
& PESQ:  & 2.61 & 3.23 & 3.29 & 3.29 & 3.35 & 3.21 & 3.24\\ 
\midrule 
\multirow{ 5}{*}{15 dB:} 
& STOI:  & 0.96 & \textbf{0.98} & \textbf{0.98} & \textbf{0.98} & \textbf{0.98} & \textbf{0.98} & 0.96\\ 
& ESTOI: & 0.87 & 0.93 & \textbf{0.94} & \textbf{0.94} & \textbf{0.94} & 0.93 & 0.89\\ 
& SI-SDR:& 13.96 & 19.13 & -1.00 & -12.08 & \textbf{19.98} & -1.72 & -17.04\\ 
& SDR:   & 14.02 & 19.74 & 13.71 & 11.24 & \textbf{20.94} & 15.70 & 7.52\\ 
& PESQ:  & 2.93 & 3.45 & 3.53 & 3.51 & 3.55 & 3.45 & 3.43\\ 
\midrule 
\multirow{ 5}{*}{20 dB:} 
& STOI:  & 0.98 & 0.98 & 0.98 & 0.98 & \textbf{0.99} & 0.98 & 0.96\\ 
& ESTOI: & 0.93 & 0.94 & 0.95 & \textbf{0.96} & \textbf{0.96} & 0.94 & 0.90\\ 
& SI-SDR:& 18.96 & 20.39 & -0.96 & -12.00 & \textbf{21.79} & -1.68 & -17.04\\ 
& SDR:   & 19.02 & 21.04 & 14.07 & 11.44 & \textbf{23.02} & 16.21 & 7.59\\ 
& PESQ:  & 3.27 & 3.62 & 3.73 & 3.69 & \textbf{3.74} & 3.64 & 3.56\\ 
\midrule 

		\end{tabular}}
	\end{subtable}
\end{table*}
\begin{table*}
	\caption{As Table\;\ref{tab:nt_res} but for the two unmatched pedestrian and bus noise types.}\label{tab:nt_res1}
	\begin{subtable}{.5\textwidth}
		\caption{Pedestrian Noise (unmatched)}\label{tab:nt_res_ped}
		\centering
		\setlength\tabcolsep{5pt} 
		\resizebox{0.99\columnwidth}{!}{%
			\begin{tabular}{cllcccccc}
				\midrule
				\multirow{2}{*}{SNR} & \multirow{2}{*}{Metric} & \multirow{2}{*}{Noisy} &\multicolumn{6}{c}{Processed} \\ \cmidrule(lr){4-9}
				&  &  &$\mathcal{L}_{\text{TIME-MSE}}$ & $\mathcal{L}_{\text{STOI}}$ & $\mathcal{L}_{\text{ESTOI}}$ & $\mathcal{L}_{\text{SI-SDR}}$ & $\mathcal{L}_{\text{STSA-MSE}}$  & $\mathcal{L}_{\text{PMSQE}}$  \\		
				\midrule 
				\multirow{ 5}{*}{-10 dB:} 
& STOI:  & 0.52 & 0.66 & \textbf{0.71} & 0.67 & 0.67 & 0.67 & 0.61\\ 
& ESTOI: & 0.19 & 0.38 & 0.44 & \textbf{0.45} & 0.39 & 0.38 & 0.31\\ 
& SI-SDR:& -11.04 & -0.06 & -7.53 & -17.82 & \textbf{0.38} & -8.05 & -20.99\\ 
& SDR:   & -10.31 & 1.08 & -0.90 & -1.80 & \textbf{1.53} & -0.40 & -1.91\\ 
& PESQ:  & 1.68 & 1.73 & 1.72 & 1.45 & 1.74 & \textbf{1.77} & 1.60\\ 
\midrule 
\multirow{ 5}{*}{-5 dB:} 
& STOI:  & 0.62 & 0.82 & \textbf{0.85} & 0.83 & 0.83 & 0.82 & 0.77\\ 
& ESTOI: & 0.29 & 0.59 & 0.64 & \textbf{0.66} & 0.61 & 0.59 & 0.52\\ 
& SI-SDR:& -6.04 & 5.45 & -3.73 & -14.37 & \textbf{5.86} & -4.24 & -18.59\\ 
& SDR:   & -5.75 & 6.21 & 4.19 & 3.40 & \textbf{6.60} & 4.87 & 2.14\\ 
& PESQ:  & 1.61 & 2.21 & 2.21 & 2.04 & 2.23 & \textbf{2.24} & 2.11\\ 
\midrule 
\multirow{ 5}{*}{0 dB:} 
& STOI:  & 0.73 & 0.90 & \textbf{0.92} & 0.91 & 0.91 & 0.90 & 0.87\\ 
& ESTOI: & 0.43 & 0.74 & 0.78 & \textbf{0.79} & 0.76 & 0.74 & 0.69\\ 
& SI-SDR:& -1.04 & 9.52 & -2.08 & -13.23 & \textbf{10.05} & -2.74 & -17.53\\ 
& SDR:   & -0.91 & 10.12 & 7.93 & 6.91 & \textbf{10.66} & 8.70 & 4.80\\ 
& PESQ:  & 1.82 & 2.60 & 2.62 & 2.55 & \textbf{2.66} & 2.62 & 2.54\\ 
\midrule 
\multirow{ 5}{*}{5 dB:} 
& STOI:  & 0.83 & 0.94 & \textbf{0.95} & \textbf{0.95} & \textbf{0.95} & 0.94 & 0.92\\ 
& ESTOI: & 0.58 & 0.83 & 0.86 & \textbf{0.87} & 0.85 & 0.83 & 0.79\\ 
& SI-SDR:& 3.96 & 12.82 & -1.40 & -12.62 & \textbf{13.47} & -2.10 & -17.20\\ 
& SDR:   & 4.04 & 13.32 & 10.57 & 9.05 & \textbf{14.03} & 11.54 & 6.32\\ 
& PESQ:  & 2.13 & 2.91 & 2.94 & 2.92 & \textbf{3.01} & 2.92 & 2.88\\ 
\midrule 
\multirow{ 5}{*}{10 dB:} 
& STOI:  & 0.91 & 0.96 & \textbf{0.97} & \textbf{0.97} & \textbf{0.97} & 0.96 & 0.94\\ 
& ESTOI: & 0.73 & 0.89 & \textbf{0.91} & \textbf{0.91} & \textbf{0.91} & 0.89 & 0.85\\ 
& SI-SDR:& 8.96 & 15.77 & -1.12 & -12.28 & \textbf{16.45} & -1.84 & -17.02\\ 
& SDR:   & 9.03 & 16.23 & 12.34 & 10.32 & \textbf{17.05} & 13.75 & 7.07\\ 
& PESQ:  & 2.46 & 3.17 & 3.21 & 3.20 & \textbf{3.29} & 3.16 & 3.16\\ 
\midrule 
\multirow{ 5}{*}{15 dB:} 
& STOI:  & 0.96 & 0.97 & \textbf{0.98} & \textbf{0.98} & \textbf{0.98} & 0.97 & 0.95\\ 
& ESTOI: & 0.85 & 0.92 & \textbf{0.94} & \textbf{0.94} & \textbf{0.94} & 0.92 & 0.88\\ 
& SI-SDR:& 13.96 & 18.28 & -1.01 & -12.16 & \textbf{19.08} & -1.73 & -17.00\\ 
& SDR:   & 14.02 & 18.79 & 13.40 & 10.99 & \textbf{19.83} & 15.23 & 7.39\\ 
& PESQ:  & 2.78 & 3.39 & 3.45 & 3.42 & \textbf{3.50} & 3.39 & 3.36\\ 
\midrule 
\multirow{ 5}{*}{20 dB:} 
& STOI:  & 0.98 & 0.98 & 0.98 & 0.98 & \textbf{0.99} & 0.98 & 0.96\\ 
& ESTOI: & 0.92 & 0.94 & 0.95 & 0.95 & \textbf{0.96} & 0.94 & 0.90\\ 
& SI-SDR:& 18.96 & 19.95 & -0.95 & -12.04 & \textbf{21.23} & -1.68 & -17.02\\ 
& SDR:   & 19.02 & 20.55 & 13.93 & 11.31 & \textbf{22.23} & 15.97 & 7.50\\ 
& PESQ:  & 3.10 & 3.55 & 3.66 & 3.61 & \textbf{3.67} & 3.58 & 3.48\\ 
\midrule 
  
		\end{tabular}}
	\end{subtable}
	\newcolumntype{a}{>{\columncolor{black}}c}
	\begin{subtable}{.5\textwidth}
		\caption{Bus Noise (unmatched)}\label{tab:nt_res_bus}
		\centering
		\setlength\tabcolsep{5pt} 
		\resizebox{0.99\columnwidth}{!}{%
			\begin{tabular}{cllcccccc}
				\midrule
				\multirow{2}{*}{SNR} & \multirow{2}{*}{Metric} & \multirow{2}{*}{Noisy} &\multicolumn{6}{c}{Processed} \\ \cmidrule(lr){4-9}
				&  &  &$\mathcal{L}_{\text{TIME-MSE}}$ & $\mathcal{L}_{\text{STOI}}$ & $\mathcal{L}_{\text{ESTOI}}$ & $\mathcal{L}_{\text{SI-SDR}}$ & $\mathcal{L}_{\text{STSA-MSE}}$  & $\mathcal{L}_{\text{PMSQE}}$  \\		
				\midrule 
				\multirow{ 5}{*}{-10 dB:} 
& STOI:  & 0.71 & 0.88 & \textbf{0.90} & \textbf{0.90} & 0.89 & 0.88 & 0.86\\ 
& ESTOI: & 0.39 & 0.70 & 0.75 & \textbf{0.76} & 0.73 & 0.70 & 0.66\\ 
& SI-SDR:& -11.04 & 8.49 & -2.34 & -13.72 & \textbf{9.19} & -3.24 & -16.39\\ 
& SDR:   & -10.35 & 9.29 & 7.48 & 6.40 & \textbf{10.00} & 7.60 & 3.72\\ 
& PESQ:  & 1.67 & 2.53 & 2.59 & 2.51 & \textbf{2.61} & 2.55 & 2.50\\ 
\midrule 
\multirow{ 5}{*}{-5 dB:} 
& STOI:  & 0.78 & 0.93 & \textbf{0.94} & \textbf{0.94} & \textbf{0.94} & 0.93 & 0.91\\ 
& ESTOI: & 0.49 & 0.80 & 0.83 & \textbf{0.84} & 0.82 & 0.80 & 0.76\\ 
& SI-SDR:& -6.04 & 11.57 & -1.59 & -12.92 & \textbf{12.33} & -2.46 & -16.89\\ 
& SDR:   & -5.77 & 12.34 & 10.03 & 8.62 & \textbf{13.15} & 10.47 & 5.44\\ 
& PESQ:  & 2.03 & 2.86 & 2.91 & 2.87 & \textbf{2.95} & 2.86 & 2.81\\ 
\midrule 
\multirow{ 5}{*}{0 dB:} 
& STOI:  & 0.85 & 0.95 & \textbf{0.96} & \textbf{0.96} & \textbf{0.96} & 0.95 & 0.94\\ 
& ESTOI: & 0.60 & 0.86 & 0.88 & \textbf{0.89} & 0.88 & 0.86 & 0.82\\ 
& SI-SDR:& -1.04 & 14.16 & -1.26 & -12.44 & \textbf{14.97} & -2.05 & -17.17\\ 
& SDR:   & -0.92 & 14.89 & 11.85 & 10.05 & \textbf{15.82} & 12.65 & 6.59\\ 
& PESQ:  & 2.37 & 3.13 & 3.17 & 3.16 & \textbf{3.22} & 3.11 & 3.08\\ 
\midrule 
\multirow{ 5}{*}{5 dB:} 
& STOI:  & 0.91 & \textbf{0.97} & \textbf{0.97} & \textbf{0.97} & \textbf{0.97} & \textbf{0.97} & 0.95\\ 
& ESTOI: & 0.71 & 0.90 & \textbf{0.92} & \textbf{0.92} & \textbf{0.92} & 0.90 & 0.87\\ 
& SI-SDR:& 3.96 & 16.51 & -1.11 & -12.20 & \textbf{17.35} & -1.85 & -17.18\\ 
& SDR:   & 4.04 & 17.21 & 13.06 & 10.89 & \textbf{18.27} & 14.32 & 7.26\\ 
& PESQ:  & 2.69 & 3.36 & 3.41 & 3.40 & 3.46 & 3.34 & 3.31\\ 
\midrule 
\multirow{ 5}{*}{10 dB:} 
& STOI:  & 0.95 & \textbf{0.98} & \textbf{0.98} & \textbf{0.98} & \textbf{0.98} & 0.97 & 0.96\\ 
& ESTOI: & 0.82 & 0.93 & \textbf{0.94} & \textbf{0.94} & \textbf{0.94} & 0.92 & 0.89\\ 
& SI-SDR:& 8.96 & 18.55 & -1.03 & -12.09 & \textbf{19.51} & -1.75 & -17.12\\ 
& SDR:   & 9.02 & 19.27 & 13.77 & 11.33 & \textbf{20.59} & 15.49 & 7.56\\ 
& PESQ:  & 3.00 & 3.58 & 3.63 & 3.61 & 3.66 & 3.54 & 3.50\\ 
\midrule 
\multirow{ 5}{*}{15 dB:} 
& STOI:  & 0.97 & 0.98 & 0.98 & 0.98 & \textbf{0.99} & 0.98 & 0.96\\ 
& ESTOI: & 0.90 & 0.94 & 0.95 & 0.95 & \textbf{0.96} & 0.94 & 0.91\\ 
& SI-SDR:& 13.96 & 19.98 & -0.99 & -12.02 & \textbf{21.27} & -1.70 & -17.09\\ 
& SDR:   & 14.02 & 20.71 & 14.10 & 11.51 & \textbf{22.58} & 16.14 & 7.65\\ 
& PESQ:  & 3.33 & 3.74 & 3.81 & 3.77 & \textbf{3.84} & 3.71 & 3.62\\ 
\midrule 
\multirow{ 5}{*}{20 dB:} 
& STOI:  & 0.99 & 0.98 & \textbf{0.99} & 0.98 & \textbf{0.99} & 0.98 & 0.96\\ 
& ESTOI: & 0.95 & 0.95 & 0.96 & 0.96 & \textbf{0.97} & 0.95 & 0.91\\ 
& SI-SDR:& 18.96 & 20.70 & -0.96 & -12.01 & \textbf{22.46} & -1.68 & -17.06\\ 
& SDR:   & 19.02 & 21.41 & 14.21 & 11.56 & \textbf{23.95} & 16.38 & 7.64\\ 
& PESQ:  & 3.66 & 3.80 & 3.91 & 3.87 & \textbf{3.97} & 3.81 & 3.68\\ 
\midrule 
  
		\end{tabular}}
	\end{subtable}
\end{table*}
\subsection{Loss Function vs. Performance Metric}\label{sec:exresLS}
We now turn our attention towards the speech enhancement potential of the systems trained to minimize the loss functions in question.  
Specifically, we study the speech enhancement performance in terms of STOI \cite{taal_algorithm_2011}, ESTOI \cite{jensen_algorithm_2016}, SI-SDR \cite{roux_sdr_2019}, SDR \cite{fevotte_bss_2011}, and PESQ \cite{rix_perceptual_2001} of six different time-domain FCNN-based speech enhancement systems when trained using the loss functions, training data, and noise-types presented in Sec.\,\ref{sec:sesys} and the learning rates given in Table\,\ref{tab:models}. 
The six systems have been tested using the matched noise types, SSN, BBL, CAF, and STR and the unmatched noise types, PED, and BUS, at SNRs from -10 dB to 20 dB and the systems are evaluated by their ability to improve the above-mentioned performance metrics.   
Note, in contrast to the training and validation data, a VAD has not been applied to the test data during inference. In other words, the speech enhancement systems process the test signals in their entirety, including any short natural occurring leading and trailing silent regions and speech pauses in between spoken words. This is done to simulate a realistic usage scenario, where exact knowledge about speech activity is generally not available prior to speech processing.       

In Table\,\ref{tab:nt_res} we present scores by the above-mentioned performance metrics partitioned into loss functions horizontally and SNR vertically.
The largest performance score for each metric and SNR is highlighted in boldface. 
From Table\,\ref{tab:nt_res} it is seen that all systems in general are able to improve all performance metrics with respect to the performance scores of the noisy unprocessed signals. An exception occurs for the SI-SDR and SDR metrics which, under some circumstances, can fail completely as previously discussed (Sec.\,\ref{sec:exresSI}) and it is important to emphasize that these systems, despite the occasionally very low SDR and SI-SDR scores, still successfully enhance the speech signals in terms of perception. This is also supported by the STOI, ESTOI, and PESQ performance metrics. 
In other words, systems trained with the six loss functions seem to be successful in terms of their ability to attenuate the noise and enhance the speech signal.       
More interestingly, although not surprising, it is seen that systems trained using $\mathcal{L}_{\text{STOI}}$, $\mathcal{L}_{\text{ESTOI}}$, and $\mathcal{L}_{\text{SI-SDR}}$ also achieve the maximum STOI, ESTOI, and SI-SDR scores, respectively. 
Somewhat surprising is it to see that systems trained to minimize $\mathcal{L}_{\text{PMSQE}}$ do not achieve the maximum PESQ score, despite the fact that $\mathcal{L}_{\text{PMSQE}}$ is designed to resemble PESQ and we see a monotonic relationship between the two functions in Fig.\,\ref{fig:pesq_vs_pmsqe}. Instead, it is seen from Table\,\ref{tab:nt_res} that systems trained to minimize $\mathcal{L}_{\text{SI-SDR}}$ generally achieve the maximum PESQ score. 
In fact, systems trained to minimize $\mathcal{L}_{\text{SI-SDR}}$ seem to perform well in general as they generally achieve large improvements across all performance metrics and often perform on par with systems trained to minimize $\mathcal{L}_{\text{STOI}}$ and $\mathcal{L}_{\text{ESTOI}}$, which are fundamentally different loss functions compared to $\mathcal{L}_{\text{SI-SDR}}$.   

In Table\,\ref{tab:nt_res1} we present performance scores achieved by the systems from Table\,\ref{tab:nt_res} but in unseen noise type conditions, using the pedestrian and bus noise types.  
From Table\,\ref{tab:nt_res1} it is seen, similarly to Table\,\ref{tab:nt_res}, that systems trained using $\mathcal{L}_{\text{STOI}}$, $\mathcal{L}_{\text{ESTOI}}$, and $\mathcal{L}_{\text{SI-SDR}}$ also achieve the maximum STOI, ESTOI, and SI-SDR scores, respectively. 
It is also seen that systems trained to minimize $\mathcal{L}_{\text{SI-SDR}}$ generally achieve the maximum, or close to the maximum, performance scores and also achieve larger PESQ scores than the systems trained to minimize $\mathcal{L}_{\text{PMSQE}}$. In other words, the behavior observed in Table\,\ref{tab:nt_res} where the systems were tested using matched noise types also seem to hold for unmatched noise types. 

Finally, from Table\,\ref{tab:nt_res} and Table\,\ref{tab:nt_res1} we can conclude that if the goal is to maximize a specific performance metric, gains can in general be achieved by training systems to minimize a loss function designed specifically to resemble that particular performance metric. 
For example, if the goal is to maximize ESTOI, the largest ESTOI scores are achieved by training systems that minimize the $\mathcal{L}_{\text{ESTOI}}$ loss function. 
However, if the goal is to perform good in general across a wide range of performance metrics, a loss function like $\mathcal{L}_{\text{SI-SDR}}$ seems to be a good candidate as systems trained to minimize $\mathcal{L}_{\text{SI-SDR}}$ achieve high improvements over a range of performance metrics.                
Also, and more importantly, these findings seem to be generally valid over a wide range of SNRs, unseen male and female speakers, as well as matched and unmatched noise types.

\section{Conclusion}\label{sec:con}
In this paper the speech enhancement potential of six state-of-the-art loss functions for time-domain deep neural network-based monaural speech enhancement have been investigated. 
Specifically, we have conducted multiple experimental studies using speech enhancement systems based on time-domain convolutional neural networks and studied the impact the loss functions have on the performance of those systems, when they are evaluated using five commonly used performance metrics for monaural speech enhancement algorithms.
The goal of the study is to establish if, and to what extent, a loss function designed specifically to resemble a certain performance metric is advantageous compared to standard loss functions such as the time-domain mean-square error\,(MSE) loss function or the short-time spectral amplitude\,(STSA)-MSE, whose strongest justification is mathematical convenience.
In addition to the classical loss functions based on time-domain MSE and STSA-MSE, we have studied a loss function based on scale-invariant signal to distortion ratio\,(SDR), as well as two loss functions based on two often used speech intelligibility predictors, namely the short-time objective intelligibility\,(STOI), and the Extended-STOI\,(ESTOI). Lastly, we have studied a loss function based on perceptual evaluation of speech quality\,(PESQ), which is a commonly used speech quality predictor. 
In general, we found that all six loss functions are good candidates for monaural speech enhancement systems as they all managed to improve the performance metrics employed with respect to the performance scores of noisy unprocessed speech signals. 
More importantly, we found that these results were generally valid across a wide range of SNRs, unseen male and female speakers, as well as matched and unmatched noise types. 
However, we also found that if the goal is to perform optimally with respect to a specific performance metric, it is generally advantageous to optimize with respect to a loss function designed specifically to resemble that particular loss function. This is particularly interesting for loss functions based on STOI and ESTOI as these performance metrics predict speech intelligibility, a metric many speech enhancement algorithms attempt to maximize without explicitly being designed to do so.
Furthermore, we found that the learning rate used when training systems to minimize a particular loss function can have a critical impact on the performance of such systems; it is paramount that the optimal learning rate is identified for each loss function, as a sub-optimal learning rate can lead to sub-optimal results and erroneous conclusions, when systems trained to optimize different loss functions are compared. Despite its obvious importance, this is a consideration that has been generally absent in the academic literature. 
Additionally, we found that waveform matching performance metrics such as SDR and SI-SDR, despite achieving good general performance, must be used with caution, when they are used in combination with speech enhancement systems with the capability of modifying the phase of the processed signals such as time-domain FCNN-based speech enhancement systems.
In particular, SDR and SI-SDR may severely under-estimate the performance of systems that are trained using loss functions that do not penalize time-shifts.   
We observed on multiple occasions that both SDR and SI-SDR failed completely, when the reference signal and the processed signal were not perfectly aligned.

In conclusion, we found that a loss function based on SI-SDR achieves good general performance across a range of popular speech enhancement evaluation metrics, which suggests that SI-SDR is a good candidate as a general-purpose loss function for supervised monaural time-domain speech enhancement.

\section*{Acknowledgment}
We would like to thank Juan M. Martín-Doñas for valuable insight and discussions regarding the implementation of the PMSQE loss function.

\bibliographystyle{bib/IEEEtran}
\bibliography{bib/mybib}

\begin{thebibliography}{10}
\providecommand{\url}[1]{#1}
\csname url@samestyle\endcsname
\providecommand{\newblock}{\relax}
\providecommand{\bibinfo}[2]{#2}
\providecommand{\BIBentrySTDinterwordspacing}{\spaceskip=0pt\relax}
\providecommand{\BIBentryALTinterwordstretchfactor}{4}
\providecommand{\BIBentryALTinterwordspacing}{\spaceskip=\fontdimen2\font plus
\BIBentryALTinterwordstretchfactor\fontdimen3\font minus
  \fontdimen4\font\relax}
\providecommand{\BIBforeignlanguage}[2]{{%
\expandafter\ifx\csname l@#1\endcsname\relax
\typeout{** WARNING: IEEEtran.bst: No hyphenation pattern has been}%
\typeout{** loaded for the language `#1'. Using the pattern for}%
\typeout{** the default language instead.}%
\else
\language=\csname l@#1\endcsname
\fi
#2}}
\providecommand{\BIBdecl}{\relax}
\BIBdecl

\bibitem{kim_algorithm_2009}
G.~Kim \emph{et~al.}, ``An algorithm that improves speech intelligibility in
  noise for normal-hearing listeners,'' \emph{The Journal of the Acoustical
  Society of America}, vol. 126, no.~3, pp. 1486--1494, 2009.

\bibitem{han_classification_2012}
K.~Han and D.~Wang, ``\BIBforeignlanguage{eng}{A classification based approach
  to speech segregation},'' \emph{\BIBforeignlanguage{eng}{The Journal of the
  Acoustical Society of America}}, vol. 132, no.~5, pp. 3475--3483, 2012.

\bibitem{wang_towards_2013}
Y.~Wang and D.~Wang, ``Towards {Scaling} {Up} {Classification}-{Based} {Speech}
  {Separation},'' \emph{IEEE/ACM Transactions on Audio, Speech, and Language
  Processing}, vol.~21, no.~7, pp. 1381--1390, 2013.

\bibitem{xu_experimental_2014}
Y.~Xu \emph{et~al.}, ``An {Experimental} {Study} on {Speech} {Enhancement}
  {Based} on {Deep} {Neural} {Networks},'' \emph{IEEE Signal Processing
  Letters}, vol.~21, no.~1, pp. 65--68, 2014.

\bibitem{weninger_single-channel_2014}
F.~Weninger, F.~Eyben, and B.~Schuller, ``Single-channel speech separation with
  memory-enhanced recurrent neural networks,'' in \emph{Proc. {ICASSP}}, 2014,
  pp. 3709--3713.

\bibitem{healy_algorithm_2015}
E.~W. Healy \emph{et~al.}, ``An algorithm to increase speech intelligibility
  for hearing-impaired listeners in novel segments of the same noise type,''
  \emph{The Journal of the Acoustical Society of America}, vol. 138, no.~3, pp.
  1660--1669, 2015.

\bibitem{chen_large-scale_2016}
J.~Chen \emph{et~al.}, ``Large-scale training to increase speech
  intelligibility for hearing-impaired listeners in novel noises,'' \emph{The
  Journal of the Acoustical Society of America}, vol. 139, no.~5, pp.
  2604--2612, 2016.

\bibitem{erdogan_deep_2017}
H.~Erdogan \emph{et~al.}, ``\BIBforeignlanguage{en}{Deep {Recurrent} {Networks}
  for {Separation} and {Recognition} of {Single}-{Channel} {Speech} in
  {Nonstationary} {Background} {Audio}},'' in \emph{\BIBforeignlanguage{en}{New
  {Era} for {Robust} {Speech} {Recognition}}}.\hskip 1em plus 0.5em minus
  0.4em\relax Springer, 2017, pp. 165--186.

\bibitem{kolbaek_speech_2017}
M.~Kolbæk, Z.~H. Tan, and J.~Jensen, ``Speech {Intelligibility} {Potential} of
  {General} and {Specialized} {Deep} {Neural} {Network} {Based} {Speech}
  {Enhancement} {Systems},'' \emph{IEEE/ACM Transactions on Audio, Speech, and
  Language Processing}, vol.~25, no.~1, pp. 153--167, 2017.

\bibitem{kolbaek_relationship_2019}
M.~Kolbæk, Z.~Tan, and J.~Jensen, ``On the {Relationship} {Between}
  {Short}-{Time} {Objective} {Intelligibility} and {Short}-{Time}
  {Spectral}-{Amplitude} {Mean}-{Square} {Error} for {Speech} {Enhancement},''
  \emph{IEEE/ACM Transactions on Audio, Speech, and Language Processing},
  vol.~27, no.~2, pp. 283--295, 2019.

\bibitem{wang_supervised_2018}
D.~Wang and J.~Chen, ``Supervised {Speech} {Separation} {Based} on {Deep}
  {Learning}: {An} {Overview},'' \emph{IEEE/ACM Transactions on Audio, Speech,
  and Language Processing}, vol.~26, no.~10, pp. 1702--1726, 2018.

\bibitem{kolbaek_single-microphone_2018}
\BIBentryALTinterwordspacing
M.~Kolbæk, ``\BIBforeignlanguage{English}{Single-{Microphone} {Speech}
  {Enhancement} and {Separation} {Using} {Deep} {Learning}},'' Ph.D.
  dissertation, Aalborg Universitetsforlag, 2018. [Online]. Available:
  \url{kolbaek-phd.aau.dk}
\BIBentrySTDinterwordspacing

\bibitem{healy_algorithm_2017}
E.~W. Healy \emph{et~al.}, ``An algorithm to increase intelligibility for
  hearing-impaired listeners in the presence of a competing talker,'' \emph{The
  Journal of the Acoustical Society of America}, vol. 141, no.~6, pp.
  4230--4239, 2017.

\bibitem{bolner_speech_2016}
F.~Bolner \emph{et~al.}, ``Speech enhancement based on neural networks applied
  to cochlear implant coding strategies,'' in \emph{Proc. {ICASSP}}, 2016, pp.
  6520--6524.

\bibitem{monaghan_auditory_2017}
J.~J.~M. Monaghan \emph{et~al.}, ``Auditory inspired machine learning
  techniques can improve speech intelligibility and quality for
  hearing-impaired listeners,'' \emph{The Journal of the Acoustical Society of
  America}, vol. 141, no.~3, pp. 1985--1998, 2017.

\bibitem{goehring_speech_2017}
T.~Goehring \emph{et~al.}, ``Speech enhancement based on neural networks
  improves speech intelligibility in noise for cochlear implant users,''
  \emph{Hearing Research}, vol. 344, pp. 183--194, 2017.

\bibitem{lai_deep_2017}
Y.~H. Lai \emph{et~al.}, ``A {Deep} {Denoising} {Autoencoder} {Approach} to
  {Improving} the {Intelligibility} of {Vocoded} {Speech} in {Cochlear}
  {Implant} {Simulation},'' \emph{IEEE Transactions on Biomedical Engineering},
  vol.~64, no.~7, pp. 1568--1578, 2017.

\bibitem{lai_deep_2018}
Y.-H. Lai \emph{et~al.}, ``Deep {Learning}-{Based} {Noise} {Reduction}
  {Approach} to {Improve} {Speech} {Intelligibility} for {Cochlear} {Implant}
  {Recipients},'' \emph{Ear and Hearing}, vol.~39, no.~4, pp. 795--809, 2018.

\bibitem{healy_deep_2019}
\BIBentryALTinterwordspacing
E.~W. Healy \emph{et~al.}, ``A deep learning algorithm to increase
  intelligibility for hearing-impaired listeners in the presence of a competing
  talker and reverberation,'' \emph{The Journal of the Acoustical Society of
  America}, vol. 145, no.~3, pp. 1378--1388, Mar. 2019. [Online]. Available:
  \url{https://asa.scitation.org/doi/full/10.1121/1.5093547}
\BIBentrySTDinterwordspacing

\bibitem{roux_phasebook:_2019}
J.~L. Roux \emph{et~al.}, ``The {Phasebook}: {Building} {Complex} {Masks} via
  {Discrete} {Representations} for {Source} {Separation},'' in \emph{{ICASSP}
  2019 - 2019 {IEEE} {International} {Conference} on {Acoustics}, {Speech} and
  {Signal} {Processing} ({ICASSP})}, May 2019, pp. 66--70.

\bibitem{wang_deep_2019}
Z.~Wang, K.~Tan, and D.~Wang, ``Deep {Learning} {Based} {Phase}
  {Reconstruction} for {Speaker} {Separation}: {A} {Trigonometric}
  {Perspective},'' in \emph{Proc. {ICASSP}}, 2019, pp. 71--75.

\bibitem{wang_end--end_2018}
Z.-Q. Wang \emph{et~al.}, ``End-to-{End} {Speech} {Separation} with {Unfolded}
  {Iterative} {Phase} {Reconstruction},'' in \emph{Proc. {Interspeech}}, 2018,
  pp. 2708--2712.

\bibitem{pandey_new_2018}
A.~Pandey and D.~Wang, ``A {New} {Framework} for {Supervised} {Speech}
  {Enhancement} in the {Time} {Domain},'' in \emph{Proc. {Interspeech}}, 2018,
  pp. 1136--1140.

\bibitem{pandey_new_2019}
------, ``A {New} {Framework} for {CNN}-{Based} {Speech} {Enhancement} in the
  {Time} {Domain},'' \emph{IEEE/ACM Transactions on Audio, Speech, and Language
  Processing}, vol.~27, no.~7, pp. 1179--1188, 2019.

\bibitem{fu_end--end_2018}
S.~W. Fu \emph{et~al.}, ``End-to-{End} {Waveform} {Utterance} {Enhancement} for
  {Direct} {Evaluation} {Metrics} {Optimization} by {Fully} {Convolutional}
  {Neural} {Networks},'' \emph{IEEE/ACM Transactions on Audio, Speech, and
  Language Processing}, vol.~26, no.~9, pp. 570 -- 1584, 2018.

\bibitem{park_fully_2017}
S.~R. Park and J.~Lee, ``A {Fully} {Convolutional} {Neural} {Network} for
  {Speech} {Enhancement},'' in \emph{Proc. {Interspeech}}, 2017, pp.
  1993--1997.

\bibitem{fu_raw_2017}
S.~W. Fu \emph{et~al.}, ``Raw waveform-based speech enhancement by fully
  convolutional networks,'' in \emph{Proc. {APSIPA}}, 2017, pp. 6--12.

\bibitem{pandey_tcnn:_2019}
A.~Pandey and D.~Wang, ``{TCNN}: {Temporal} {Convolutional} {Neural} {Network}
  for {Real}-time {Speech} {Enhancement} in the {Time} {Domain},'' in
  \emph{Proc. {ICASSP}}, 2019, pp. 6875--6879.

\bibitem{grzywalski_using_2019}
T.~Grzywalski and S.~Drgas, ``Using {Recurrences} in {Time} and {Frequency}
  within {U}-net {Architecture} for {Speech} {Enhancement},'' in \emph{Proc.
  {ICASSP}}, 2019, pp. 6970--6974.

\bibitem{tan_real-time_2019}
K.~Tan, X.~Zhang, and D.~Wang, ``Real-time {Speech} {Enhancement} {Using} an
  {Efficient} {Convolutional} {Recurrent} {Network} for {Dual}-microphone
  {Mobile} {Phones} in {Close}-talk {Scenarios},'' in \emph{Proc. {ICASSP}},
  2019, pp. 5751--5755.

\bibitem{ephraim_speech_1984}
Y.~Ephraim and D.~Malah, ``Speech enhancement using a minimum-mean square error
  short-time spectral amplitude estimator,'' \emph{IEEE Transactions on
  Acoustics, Speech, and Signal Processing}, vol.~32, no.~6, pp. 1109--1121,
  1984.

\bibitem{taal_algorithm_2011}
C.~H. Taal \emph{et~al.}, ``An {Algorithm} for {Intelligibility} {Prediction}
  of {Time}-{Frequency} {Weighted} {Noisy} {Speech},'' \emph{IEEE/ACM
  Transactions on Audio, Speech, and Language Processing}, vol.~19, no.~7, pp.
  2125--2136, 2011.

\bibitem{jensen_algorithm_2016}
J.~Jensen and C.~H. Taal, ``An {Algorithm} for {Predicting} the
  {Intelligibility} of {Speech} {Masked} by {Modulated} {Noise} {Maskers},''
  \emph{IEEE/ACM Transactions on Audio, Speech, and Language Processing},
  vol.~24, no.~11, pp. 2009--2022, 2016.

\bibitem{roux_sdr_2019}
J.~L. Roux \emph{et~al.}, ``{SDR} – {Half}-baked or {Well} {Done}?'' in
  \emph{{ICASSP} 2019}, 2019, pp. 626--630.

\bibitem{martin-donas_deep_2018}
J.~M. Martín-Doñas \emph{et~al.}, ``A {Deep} {Learning} {Loss} {Function}
  {Based} on the {Perceptual} {Evaluation} of the {Speech} {Quality},''
  \emph{IEEE Signal Processing Letters}, vol.~25, no.~11, pp. 1680--1684, 2018.

\bibitem{venkataramani_performance_2018}
S.~Venkataramani, R.~Higa, and P.~Smaragdis, ``Performance {Based} {Cost}
  {Functions} for {End}-to-{End} {Speech} {Separation},'' \emph{Proc. APSIPA},
  pp. 350--355, 2018.

\bibitem{venkataramani_end--end_2017}
S.~Venkataramani, J.~Casebeer, and P.~Smaragdis, ``End-to-end {Source}
  {Separation} with {Adaptive} {Front}-{Ends},'' in \emph{Proc. {NIPS}
  {Machine} {Learning} for {Audio} {Signal} {Processing} {Workshop}}, 2017.

\bibitem{zhao_perceptually_2018}
Y.~Zhao \emph{et~al.}, ``Perceptually {Guided} {Speech} {Enhancement} using
  {Deep} {Neural} {Networks},'' in \emph{Proc. {ICASSP}}, 2018, pp. 5074--5078.

\bibitem{zhang_training_2018}
H.~Zhang, X.~Zhang, and G.~Gao, ``Training {Supervised} {Speech} {Separation}
  {System} to {Improve} {STOI} and {PESQ} {Directly},'' in \emph{Proc.
  {ICASSP}}, 2018, pp. 5374--5378.

\bibitem{bahmaninezhad_comprehensive_2019}
F.~Bahmaninezhad \emph{et~al.}, ``A {Comprehensive} {Study} of {Speech}
  {Separation}: {Spectrogram} vs {Waveform} {Separation},'' in \emph{Proc.
  {Interspeech}}, 2019, pp. 4574--4578.

\bibitem{luo_conv-tasnet:_2019}
Y.~Luo and N.~Mesgarani, ``Conv-{TasNet}: {Surpassing} {Ideal}
  {Time}–{Frequency} {Magnitude} {Masking} for {Speech} {Separation},''
  \emph{IEEE/ACM Transactions on Audio, Speech, and Language Processing},
  vol.~27, no.~8, pp. 1256--1266, May 2019.

\bibitem{luo_tasnet:_2018-1}
------, ``{TaSNet}: {Time}-{Domain} {Audio} {Separation} {Network} for
  {Real}-{Time}, {Single}-{Channel} {Speech} {Separation},'' in \emph{Proc.
  {ICASSP}}, 2018, pp. 696--700.

\bibitem{kolbaek_monaural_2018-1}
M.~Kolbæk, Z.-H. Tan, and J.~Jensen, ``Monaural {Speech} {Enhancement} using
  {Deep} {Neural} {Networks} by {Maximizing} a {Short}-{Time} {Objective}
  {Intelligibility} {Measure},'' in \emph{Proc. {ICASSP}}, 2018, pp. 5059 --
  5063.

\bibitem{kolbaek_multi-talker_2017-1}
M.~Kolbæk \emph{et~al.}, ``Multi-talker {Speech} {Separation} {With}
  {Utterance}-{Level} {Permutation} {Invariant} {Training} of {Deep}
  {Recurrent} {Neural} {Networks},'' \emph{IEEE/ACM Transactions on Audio,
  Speech, and Language Processing}, vol.~25, no.~10, pp. 1901--1913, Jul. 2017.

\bibitem{wang_training_2014}
Y.~Wang, A.~Narayanan, and D.~Wang, ``On {Training} {Targets} for {Supervised}
  {Speech} {Separation},'' \emph{IEEE/ACM Transactions on Audio, Speech, and
  Language Processing}, vol.~22, no.~12, pp. 1849--1858, 2014.

\bibitem{naithani_deep_2018-1}
G.~Naithani \emph{et~al.}, ``Deep {Neural} {Network} {Based} {Speech}
  {Separation} {Optimizing} an {Objective} {Estimator} of {Intelligibility} for
  {Low} {Latency} {Applications},'' in \emph{Proc. {IWAENC}}, 2018, pp.
  386--390.

\bibitem{tan_gated_2019}
K.~Tan, J.~Chen, and D.~Wang, ``Gated {Residual} {Networks} {With} {Dilated}
  {Convolutions} for {Monaural} {Speech} {Enhancement},'' \emph{IEEE/ACM
  Transactions on Audio, Speech, and Language Processing}, vol.~27, no.~1, pp.
  189--198, 2019.

\bibitem{kingma_adam:_2015}
D.~P. Kingma and J.~Ba, ``Adam: {A} {Method} for {Stochastic} {Optimization},''
  in \emph{Proc. {ICLR} ({arXiv}:1412.6980)}, 2015.

\bibitem{baby_sergan:_2019}
D.~Baby and S.~Verhulst, ``Sergan: {Speech} {Enhancement} {Using}
  {Relativistic} {Generative} {Adversarial} {Networks} with {Gradient}
  {Penalty},'' in \emph{{ICASSP} 2019 - 2019 {IEEE} {International}
  {Conference} on {Acoustics}, {Speech} and {Signal} {Processing} ({ICASSP})},
  May 2019, pp. 106--110.

\bibitem{pascual_segan:_2017}
S.~Pascual, A.~Bonafonte, and J.~Serrà, ``{SEGAN}: {Speech} {Enhancement}
  {Generative} {Adversarial} {Network},'' in \emph{Proc. {INTERSPEECH}}, 2017,
  pp. 3642--3646.

\bibitem{ernst_speech_2018}
O.~Ernst \emph{et~al.}, ``Speech {Dereverberation} {Using} {Fully}
  {Convolutional} {Networks},'' in \emph{Proc. {EUSIPCO}}, 2018, pp. 390--394.

\bibitem{loizou_speech_2013}
P.~C. Loizou, \emph{Speech {Enhancement}: {Theory} and {Practice}}.\hskip 1em
  plus 0.5em minus 0.4em\relax CRC Press, 2013.

\bibitem{bishop_pattern_2006}
C.~M. Bishop, \emph{Pattern {Recognition} and {Machine} {Learning}}.\hskip 1em
  plus 0.5em minus 0.4em\relax Springer, 2006.

\bibitem{goodfellow_deep_2016}
I.~Goodfellow, Y.~Bengio, and A.~Courville, \emph{Deep {Learning}}.\hskip 1em
  plus 0.5em minus 0.4em\relax MIT Press, 2016.

\bibitem{hendriks_dft-domain_2013}
R.~C. Hendriks, T.~Gerkmann, and J.~Jensen, ``{DFT}-{Domain} {Based}
  {Single}-{Microphone} {Noise} {Reduction} for {Speech} {Enhancement}: {A}
  {Survey} of the {State} of the {Art},'' \emph{Synthesis Lectures on Speech
  and Audio Processing}, vol.~9, no.~1, pp. 1--80, 2013.

\bibitem{jensen_speech_2014}
J.~Jensen and C.~H. Taal, ``Speech {Intelligibility} {Prediction} {Based} on
  {Mutual} {Information},'' \emph{IEEE/ACM Transactions on Audio, Speech, and
  Language Processing}, vol.~22, no.~2, pp. 430--440, 2014.

\bibitem{fevotte_bss_2011}
C.~Févotte, R.~Gribonval, and E.~Vincent, ``{BSS} {EVAL} {Toolbox} {User}
  {Guide} – {Revision} 2.0,'' IRISA, Tech. Rep. inria-00564760, 2011.

\bibitem{moore_introduction_2013}
B.~Moore, \emph{An {Introduction} to the {Psychology} of {Hearing}}.\hskip 1em
  plus 0.5em minus 0.4em\relax Brill, 2013.

\bibitem{rix_perceptual_2001}
A.~W. Rix \emph{et~al.}, ``Perceptual evaluation of speech quality ({PESQ})-a
  new method for speech quality assessment of telephone networks and codecs,''
  in \emph{Proc. {ICASSP}}, vol.~2, 2001, pp. 749--752.

\bibitem{noauthor_international_2003}
\BIBentryALTinterwordspacing
``International {Telecommunication} {Union} - {Recommendation}
  {P}.862.1 : {Mapping} function for transforming {P}.862 raw result scores
  to {MOS}-{LQO},'' 2003. [Online]. Available:
  \url{https://www.itu.int/rec/T-REC-P.862.1-200311-I/en}
\BIBentrySTDinterwordspacing

\bibitem{garofolo_csr-i_1993}
J.~S. Garofolo \emph{et~al.}, ``{CSR}-{I} ({WSJ0}) {Complete} {LDC93S6A},''
  1993, philadelphia: Linguistic Data Consortium.

\bibitem{garofolo_timit_1993}
------, ``{TIMIT} {Acoustic}-{Phonetic} {Continuous} {Speech} {Corpus}
  {LDC93S1},'' 1993, linguistic Data Consortium.

\bibitem{barker_third_2015}
J.~Barker \emph{et~al.}, ``The third ‘{CHiME}’ speech separation and
  recognition challenge: {Dataset}, task and baselines,'' in \emph{Proc.
  {ASRU}}, 2015, pp. 504--511.

\bibitem{itu_rec._2011}
ITU, ``Rec. {P}.56 : {Objective} measurement of active speech level,'' 2011,
  https://www.itu.int/rec/T-REC-P.56/.

\bibitem{ronneberger_u-net:_2015}
O.~Ronneberger, P.~Fischer, and T.~Brox, ``U-{Net}: {Convolutional} {Networks}
  for {Biomedical} {Image} {Segmentation},'' in \emph{Proc. {MICCAI}}, N.~Navab
  \emph{et~al.}, Eds., 2015, pp. 234--241.

\bibitem{he_delving_2015}
K.~He \emph{et~al.}, ``Delving {Deep} into {Rectifiers}: {Surpassing}
  {Human}-{Level} {Performance} on {ImageNet} {Classification},'' in
  \emph{Proc. {ICCV}}, 2015, pp. 1026--1034.

\bibitem{liu_variance_2020}
L.~Liu \emph{et~al.}, ``On the {Variance} of the {Adaptive} {Learning} {Rate}
  and {Beyond},'' in \emph{Proc. {ICLR}}, 2020.

\bibitem{stone_tolerable_1999}
M.~A. Stone and B.~C. Moore, ``Tolerable hearing aid delays. {I}. {Estimation}
  of limits imposed by the auditory path alone using simulated hearing
  losses,'' \emph{Ear and Hearing}, vol.~20, no.~3, pp. 182--192, 1999.

\bibitem{bramslow_preferred_2010}
L.~Bramsløw, ``Preferred signal path delay and high-pass cut-off in open
  fittings,'' \emph{International Journal of Audiology}, vol.~49, no.~9, pp.
  634--644, 2010.

\end{thebibliography}
%
%
%
%
\begin{IEEEbiography}[{\includegraphics[width=1in,height=1.25in,clip,keepaspectratio]{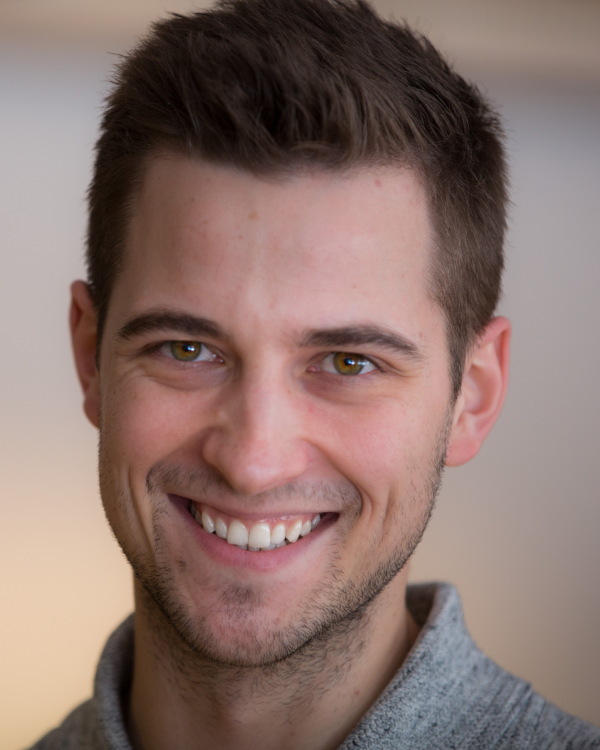}}]%
	{Morten Kolbæk}
	received the B.Eng. degree in electronic design at Aarhus University, in 2013 and the M.Sc. in signal processing and computing from Aalborg University, Denmark, in 2015. He received the PhD degree from Aalborg University, Denmark, in 2018 for the thesis entitled Single-Microphone Speech Enhancement and Separation Using Deep Learning (\url{kolbaek-phd.aau.dk}). He is currently a post-doctoral researcher at the section for Signal and Information Processing at the Department of Electronic Systems, Aalborg University, Denmark. His research interests include speech enhancement and separation, deep learning, and intelligibility improvement of noisy speech. 
\end{IEEEbiography}

\vspace*{-2\baselineskip}
\begin{IEEEbiography}
	[{\includegraphics[width=1in,height=1.25in,clip,keepaspectratio]{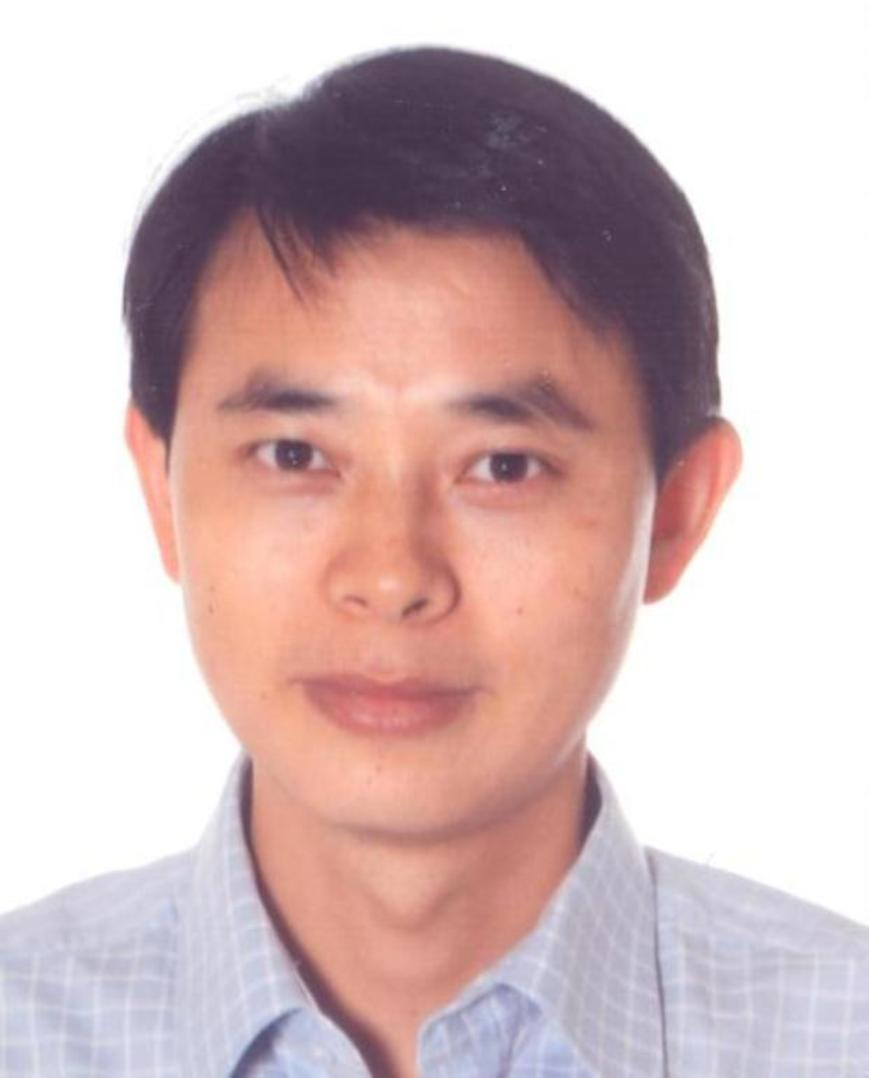}}]
	{Zheng-Hua Tan}
	(M'00--SM'06) received the B.Sc. and M.Sc. degrees in electrical engineering from Hunan University, Changsha, China, in 1990 and 1996, respectively, and the Ph.D. degree in electronic engineering from Shanghai Jiao Tong University, Shanghai, China, in 1999. He is a Professor and a Co-Head of the Centre for Acoustic Signal Processing Research (CASPR) at Aalborg University, Aalborg, Denmark. He was a Visiting Scientist at the Computer Science and Artificial Intelligence Laboratory, MIT, Cambridge, USA, an Associate Professor at Shanghai Jiao Tong University, and a postdoctoral fellow at KAIST, Daejeon, Korea. His research interests include machine learning, deep learning, pattern recognition, speech and speaker recognition, noise-robust speech processing, multimodal signal processing, and social robotics. He is the vice chair of the IEEE Signal Processing Society Machine Learning for Signal Processing Technical Committee (MLSP TC). He is an Associate Editor for IEEE/ACM Transactions on Audio, Speech and Language Processing, an Editorial Board Member for Computer Speech and Language and was a Guest Editor for the IEEE Journal of Selected Topics in Signal Processing and Neurocomputing. He was the General Chair for IEEE MLSP 2018 and a TPC co-chair for IEEE SLT 2016.
\end{IEEEbiography}

\vspace*{-2\baselineskip}
\begin{IEEEbiography}
	[{\includegraphics[width=1in,height=1.25in,clip,keepaspectratio]{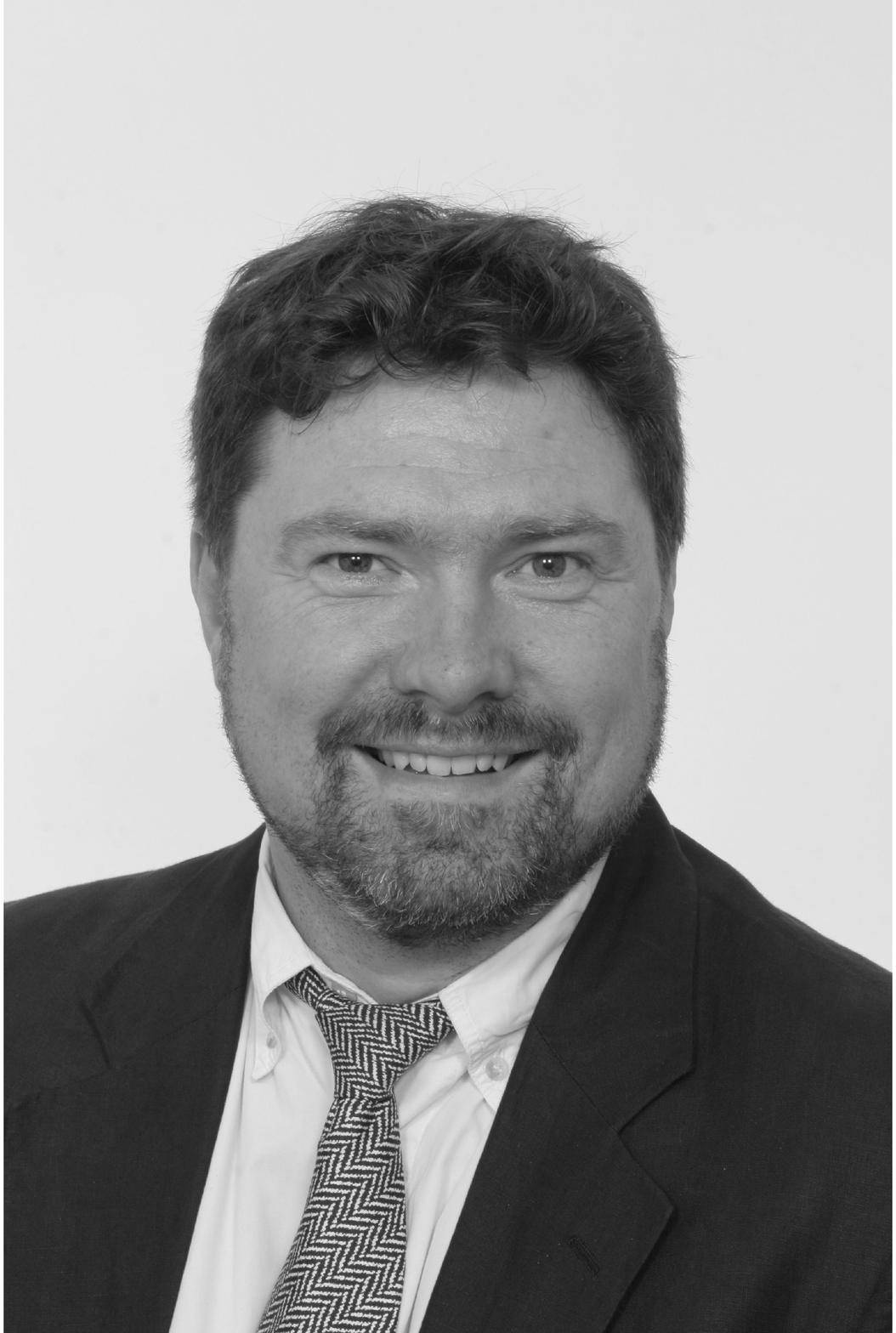}}]
	{S\o ren Holdt Jensen} (S’87–M’88–SM’00) received the M.Sc.\ degree in electrical engineering from Aalborg University (AAU), Aalborg, Denmark, in 1988, and the Ph.D.\ degree (in signal processing) from the Technical University of Denmark (DTU), Lyngby, Denmark, in 1995. He is Full Professor in Signal Processing at Aalborg University. Before joining the Department of Electronic Systems, Aalborg University, he was with the Telecommunications Laboratory of Telecom Denmark, Ltd, Taastrup (Copenhagen), Denmark; the Electronics Institute of Technical University of Denmark; the Scientific Computing Group of Danish Computing Center for Research and Education (UNI{\textbullet}C), Lyngby; the Electrical Engineering Department (ESAT-SISTA)  of Katholieke Universiteit Leuven, Leuven, Belgium; and the Center for PersonKommunikation (CPK) of Aalborg University. His current research interest are in statistical signal processing, numerical algorithms, optimization engineering, machine learning, and digital processing of acoustic, audio, communication, multimedia, and speech, signals. He is co-author of the textbook {\em Software-Defined GPS and Galileo Receiver---A Single-Frequency Approach}, Birkh{\"a}user, Boston, USA, also translated to Chinese: National Defence Industry Press, China. Prof.\ Jensen has been Associate Editor for the IEEE Transactions on Signal Processing, IEEE/ACM Transactions on Audio, Speech and Language Processing, Elsevier Signal Processing, and EURASIP Journal on Advances in Signal Processing. He is a recipient of an individual European Community Marie Curie (HCM: Human Capital and Mobility) Fellowship, former Chairman of the IEEE Denmark Section and the IEEE Denmark Section’s Signal Processing Chapter (founder and first chaiman). He is member of the Danish Academy of Technical Sciences (ATV) and has been member of the Danish Council for Independent Research (2011--2016) appointed by Danish Ministers of Science.
\end{IEEEbiography}

\vspace*{-2\baselineskip}
\begin{IEEEbiography}[{\includegraphics[width=1in,
		height=1.25in,clip,keepaspectratio]{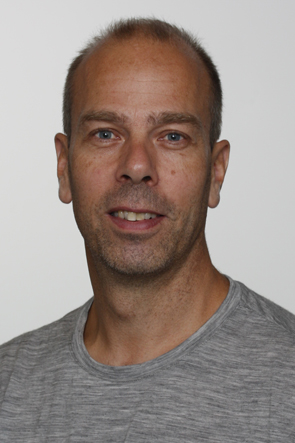}}]{Jesper Jensen} received the M.Sc. degree in electrical engineering and the Ph.D. degree in signal processing from Aalborg University, Aalborg, Denmark, in 1996 and 2000, respectively. From 1996 to 2000, he was with the Center for Person Kommunikation (CPK), Aalborg University, as a Ph.D. student and Assistant Research Professor. From 2000 to 2007, he was a Post-Doctoral Researcher and Assistant Professor with Delft University of Technology, Delft, The Netherlands, and an External Associate Professor with Aalborg University. Currently, he is a Senior Principal Scientist with Oticon A/S, Copenhagen, Denmark, where his main responsibility is scouting and development of new signal processing concepts for hearing aid applications. He is a Professor with the Section for Signal and Information Processing (SIP), Department of Electronic Systems, at Aalborg University. He is also a co-founder of the Centre for Acoustic Signal Processing Research (CASPR) at Aalborg University. His main interests are in the area of acoustic signal processing, including signal retrieval from noisy observations, coding, speech and audio modification and synthesis, intelligibility enhancement of speech signals, signal processing for hearing aid applications, and perceptual aspects of signal processing.
\end{IEEEbiography}
\end{document}